\documentclass[conference]{IEEEtran}
\IEEEoverridecommandlockouts
\usepackage{cite}
\usepackage{amsmath,amssymb,amsfonts}
\usepackage{algorithmic}
\usepackage{graphicx}
\usepackage{textcomp}
\usepackage[table,xcdraw]{xcolor}

\usepackage{cleveref}
\usepackage{tcolorbox}
\tcbuselibrary{xparse,skins,breakable}
\tcbset{%
    basestyle/.style={
        enhanced,
        breakable,
        fonttitle=\scshape
    },
    defstyle/.style={colframe=gray!25, coltitle=black, arc=0mm, left=0.8mm, right=0.8mm, top=0.8mm, bottom=0.8mm},
}
\DeclareTColorBox[auto counter, crefname={definition}{definitions}]
    {definition}{ o O{} t\label g }
    {%
        defstyle,
        IfValueTF={#1}{title={\textbf{Definition~\thetcbcounter\ | #1}}}{title=Definition~\thetcbcounter}, IfBooleanTF={#3}{label=#4}{},
        #2
    }

\usepackage{xspace}
\newcommand{\al}{\textit{et al.}\xspace}
\newcommand{\ie}{\textit{i.e.,}\xspace}

\def\BibTeX{{\rm B\kern-.05em{\sc i\kern-.025em b}\kern-.08em
    T\kern-.1667em\lower.7ex\hbox{E}\kern-.125emX}}
\begin{document}

\title{Mapping breakpoint types: an exploratory study}

\author{
\IEEEauthorblockN{Eduardo Andreetta Fontana}
\IEEEauthorblockA{
\textit{University of Quebec at Chicoutimi}\\
Saguenay - QC, Canada\\
eduardo.andreetta-fontana1@uqac.ca}
\and
\IEEEauthorblockN{Fabio Petrillo}
\IEEEauthorblockA{
\textit{University of Quebec at Chicoutimi}\\
Saguenay - QC, Canada\\
fabio@petrillo.com}
}

\maketitle

\thispagestyle{plain}
\pagestyle{plain}

\begin{abstract}
Debugging is a relevant task for finding bugs during software development, maintenance, and evolution. During  debugging, developers use modern IDE debuggers to analyze variables, step execution, and set breakpoints. 
Observing IDE debuggers, we find several breakpoint types. However, what are the breakpoint types?  
The goal of our study is to map the breakpoint types among IDEs and academic literature. 
Thus, we mapped the gray literature on the documentation of the nine main IDEs used by developers according to the three public rankings. In addition, we performed a systematic mapping of academic literature over 68 articles describing breakpoint types. Finally, we analyzed the developers understanding of the main breakpoint types through a questionnaire. We present three main contributions: (1) the mapping of breakpoint types (IDEs and literature), (2) compiled definitions of breakpoint types, (3) a breakpoint type taxonomy.
Our contributions provide the first step to organize breakpoint IDE taxonomy and lexicon, and support further debugging research.
\end{abstract}


\section{Introduction}
\label{intro}

Debugging is a relevant task for finding and resolving bugs during software development, maintenance, and evolution \cite{tanenbaum1973people}. During debugging, developers use modern interactive debuggers with Integrated Development Environments (IDEs), such as Eclipse, NetBeans, IntelliJ IDEA, and Visual Studio, to detect, locate, and correct faults in software systems \cite{petrillodevelopers2017}. Among the many features provided by IDEs, breakpoints are one of the most used features by developers to initiate a debugging session \cite{beller2017developers}. Observing IDEs documentation, we found several breakpoint types, each one has particular functionalities that could be explored by developers \cite{EclipseDoucmentation, EclipseDoucmentationNews, IntelliJDocumentation, VSDocumentation, VSCodeDocumentation}.

However, what are the breakpoint types? To the best of our knowledge, there is a lack of congruence related to breakpoint type names and descriptions in IDEs documentation and academic literature. 
We found some IDEs documentation using different names for the similar breakpoint types. For example, Eclipse \cite{EclipseDoucmentation, EclipseDoucmentationNews} and IntelliJ \cite{IntelliJDocumentation} call data-sensitive breakpoints as Watchpoint or as Field Watchpoint, but Visual Studio \cite{VSDocumentation} or Visual Studio Code \cite{VSCodeDocumentation} call them Data Breakpoint.
Furthermore, the IDEs documentation information is not clear or sufficient to support developers understanding of breakpoint types. 
For example, there are discussions in gray literature about the difference between Breakpoint, Watchpoint, Method Breakpoint \cite{DiffWatchBreakQuora, DiffWatchMethodStack, DiffBreakWatchSAP, DiffBreakWatchLessBro}, Static Breakpoint, Dynamic Breakpoint \cite{DiffStaticDynamicGoCoding}, Conditional Breakpoint, and Data Breakpoint \cite{ConditionalBreakVSMagazine, DataBreakVS2019MicrosoftDevBlog, DataBreakVS2017MicrosoftDevBlog}.

In academic literature, we found some studies that try to define or classify some breakpoint types, but there are no studies that map the breakpoint types among different IDEs and different technologies.
For instance, Petrillo \al \cite{petrillodevelopers2017, petrillo2019swarm} and Johnson \cite{johnson1982software} classify some breakpoint types as Static Breakpoint and Dynamic Breakpoint; and Keppel \cite{keppel1993fast}, Vasudevan \cite{vasudevan2009re}, and Spinellis \cite{spinellis2006debuggers} use the name Code Breakpoint, but IDEs documentation do not use these categorization names. Arya \al \cite{arya2017transition}, Spinellis \cite{spinellis2006debuggers}, Kumar \al \cite{kumar2013behave}, Copperman \al \cite{copperman1995poor}, and Zhao \cite{zhao2008million} present breakpoint types with the names of Watchpoint or Data Breakpoint at the software level, while Chew \al \cite{chew2010kivati}, Sommer \al \cite{sommer2013minerva}, and Jang \al \cite{jang2019revisiting} present the same breakpoint type names, but at the hardware level or with hardware support.

The goal of our study is to map the breakpoint types among different IDEs and academic literature. Our survey provide to developers a common lexicon to IDEs. In addition, 
our study contributes to avoid misleading breakpoint names among academic literature and IDE documentation. Furthermore, as suggested by Aniche \al \cite{aniche2021developers}, our study can support future debugging research. Finally, our study provides a discussion that establishes characteristics and the state of the art of the main existing breakpoint types and support further debugging research.

We mapped the gray literature on the documentation of the nine main IDEs used by developers according to the Stack Overflow Survey 2019 \cite{stackOverflowSurvey2019}, Google Trends 2020 \cite{googleTrends2020}, and G2 2020 \cite{g22020} rankings. In addition, we performed a systematic mapping of academic literature over 68 articles describing breakpoint types. Finally, we analyzed the developers understanding of the main breakpoint types administering a questionnaire. From the collected data, we applied The Grounded Theory method \cite{glaser2017discovery} to map and organize the definitions and characteristics of the main existing breakpoint types.

As a result, we built a breakpoint type mapping table of the main IDEs, as well as a breakpoint type mapping table for academic papers. Furthermore, we present breakpoint type definitions structured according to attributes mapped. Finally, we present a taxonomy diagram with two breakpoint type categories.

We present three main contributions: 
(1) the mapping of the existing breakpoint types in the main IDEs and academic literature,
(2) compiled definitions of breakpoint types, and
(3) a breakpoint type taxonomy.


\section{Background}
\label{sec:background}
The use of breakpoint originates in the 1940s during discovery attempts of program problems running on ENIAC (Electronic Numerical Integrator and Computer), the first digital computer in history \cite{haigh2016eniac} \cite{haigh2014engineering}. In the initial design of the ENIAC, the program flow was set by plugging cables from one unit to another. To make the program stop at a certain point, a cable was simply removed. The removal of such cable midway through the program execution was named as breakpoint. The term breakpoint was coined by Frances Elizabeth Holberton, one of the six woman programmers that worked at ENIAC \cite{tropp_holberton_1973}.

\begin{quote}
    \textit{“Well you know, the thing is we did develop the one word that's in the language today, which is `breakpoint', at that time. Because we actually did pull the wire to stop the programs so we could read the accumulators off. This was, I mean we actually broke the point, and that was where the word came from.”} - Frances Elizabeth Holberton \cite{tropp_holberton_1973}.
\end{quote}

Iterative breakpoints were initially used in the 1970s, during the era of mainframe computers, over OLIVER debugger (CICS interactive test/debug). OLIVER was a proprietary testing and debugging toolkit for interactively testing programs designed to run on IBM Customer Information Control System (CICS) on IBM System/360/370/390 architecture. It provided instruction step, conditional program breakpoint, and storage alteration features for programs written in Assembly, COBOL, and PL/I. HLL (high-level language) users were able to see and modify variables directly at a breakpoint \cite{cristobal2011ibm}.

The first use of breakpoints in IDEs (Integrated Development Environment) was observed in Turbo Pascal in the 1980s. The Turbo Pascal IDE provided several debugging facilities, including single stepping, examination and changing of variables, single breakpoints, and conditional breakpoints. In later versions, assembly language blocks could be stepped through in debugging process. The user could add breakpoints on variables and registers in an IDE window \cite{carroll1985programming}.

Currently, the use of breakpoint is an integral part of most IDEs and one of the main resources used during the debugging process. Breakpoints are also features of symbolic debuggers like GDB \cite{StallmanR.Pesch2002} and WinDbg \cite{vostokov2008windbg}, so they are part of IDE instrumentation. Some popular IDEs that use breakpoints are: Eclipse, IntelliJ, NetBeans, Visual Studio, Visual Studio Code, and others.


\section{Related Work}
\label{sec:relatedWork}


Petrillo \al \cite{petrillodevelopers2017, petrillo2019swarm}, Johnson \cite{johnson1982software}, and Chern \al \cite{Chern:2007} classify some breakpoint types as Static Breakpoint, and Dynamic Breakpoint.
However, our research extends this classification including other criteria that determine the breakpoint type groups, such as how the breakpoint is added and the effect when it is triggered.

Vasudevan \cite{vasudevan2009re} describes a generic way classification of breakpoint types through Software Breakpoint and Hardware Breakpoint categories. 
Spinellis \cite{spinellis2006debuggers} classifies software breakpoints by the term Code Breakpoint and classifies the breakpoints with related data as Data Breakpoint, with or without hardware support. Beller \al \cite{beller2018dichotomy} present categories closer to those found in IDEs, using terms such as Line Breakpoint, Exception Breakpoint, Method Breakpoint, Watchpoint, and others. 
In general, these articles describe generic categories and do not provide details or attributes of each category. Our study extends those categories, adding details, describing subcategories and relationships among them.

Moritz \al \cite{beller2018dichotomy} provide an attempt to group breakpoint types with the same nomenclature used by IDEs. Robert Wahbe \cite{wahbe1992efficient}, Zhao \al \cite{zhao2008million}, and Copperman \al \cite{copperman1995poor} provide definitions for Data Breakpoints and Watchpoints. Dupriez \al \cite{dupriez2017analysis} and Corrodi \cite{corrodi2016towards} define a new generation of debuggers, which uses the concept of context breakpoint, that is, breakpoints centered on objects already instantiated in memory.

Kumar \al \cite{kumar2013behave} introduces the Behavioral Watchpoint into Watchpoint category, as a subcategory of Software Breakpoint. 
Copperman \al \cite{copperman1995poor} and Zhao \cite{zhao2008million} present specific implementations for Watchpoint and mention that this type is a synonym for the term Data Breakpoint. Wahbe \cite{wahbe1992efficient} introduces the Data Breakpoint category to describe data-related breakpoint implementation strategies with or without hardware support. Our article extends these classifications and details Data Breakpoint and Watchpoint as synonyms within the context of data-related breakpoint.

\section{Study Design}
\label{sec:methodology}
\begin{figure}[ht]
\centering
\includegraphics[width=\linewidth]{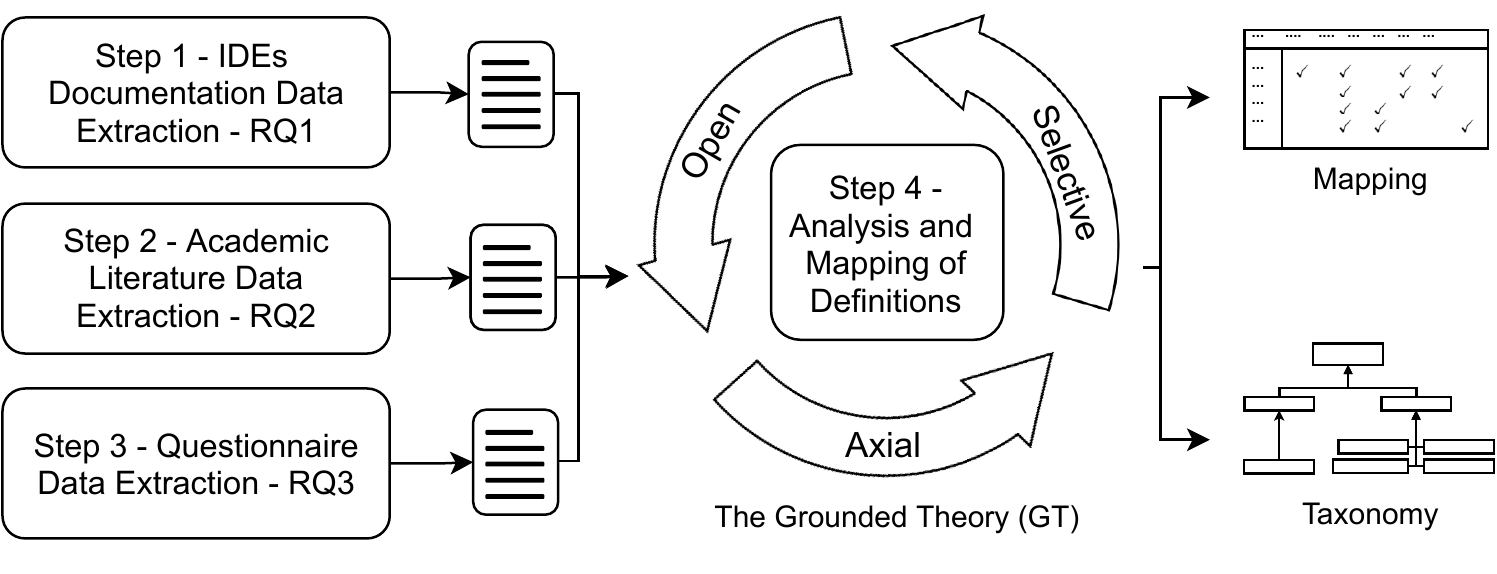}
\caption{\textbf{Protocol Steps} -- We extracted Breakpoint type definitions from three data sources, after it we processed it through The Grounded Theory method which generated as output a Mapping and a Taxonomy.}
\label{fig:figMethodologyProtocolSteps}
\end{figure}

In this section, we describe the details our study design. We performed systematic mapping of academic literature and gray literature, \ie IDEs documentation. After that, we administered a survey questionnaire. From those instruments, we analyzed and mapped breakpoint type definitions applying The Grounded Theory method \cite{glaser2017discovery}. Our protocol steps is presented in Figure \ref{fig:figMethodologyProtocolSteps}. The steps are focus on answering research questions RQ1, RQ2, and RQ3.

\subsection{Research Questions}
\label{subsec:researchQuestions}

\textbf{RQ1 - What breakpoint types are found in IDEs?} By answering this research question, we aim to map the breakpoint types used by gray literature as IDEs documentation from different IDEs and different technologies. 

\textbf{RQ2 - What breakpoint types are studied in the academic literature?} By answering this research question, we aim to map the breakpoint types investigated by academic literature.

\textbf{RQ3 - What are the breakpoint types according to practitioners?} By answering this research question, we aim to discover what breakpoint types are according to software developers.

\subsection{Scope}
\label{subsec:scope}
The scope of our research consisted of three data sources. The first data source was the official documentation of the nine main IDEs used for developing high-level software. To select IDEs, we considered the crossing of the IDE rankings websites.

The second data source was four specialized academic libraries in software engineering, in which we performed a systematic mapping, as shown in Figure \ref{fig:figArticleSelectionProcess}. From these libraries, we found the total of 931 candidate articles. Over candidate articles, we applied some filters and selections resulting in 68 articles with breakpoint type descriptions. From these 68 articles we performed the extraction of breakpoint type definitions. 

The third data source was the questionnaire administered to debugging practitioners, who are software developers. The questionnaire has 30 questions divided into three sections. We collected 29 valid responses, \ie form completed by the end.

\subsection{Protocol}
\label{subsec:Protocol}
In this subsection, we describe in detail the activities performed in each step of our study design, as shown in Figure \ref{fig:figMethodologyProtocolSteps}.

\subsubsection{Step 1 - IDEs Documentation Data Extraction - RQ1}
\label{subsubsec:Step1IDEDocumentationDataExtractionRQ1}

\begin{figure}[ht]
\centering
\includegraphics[width=\linewidth]{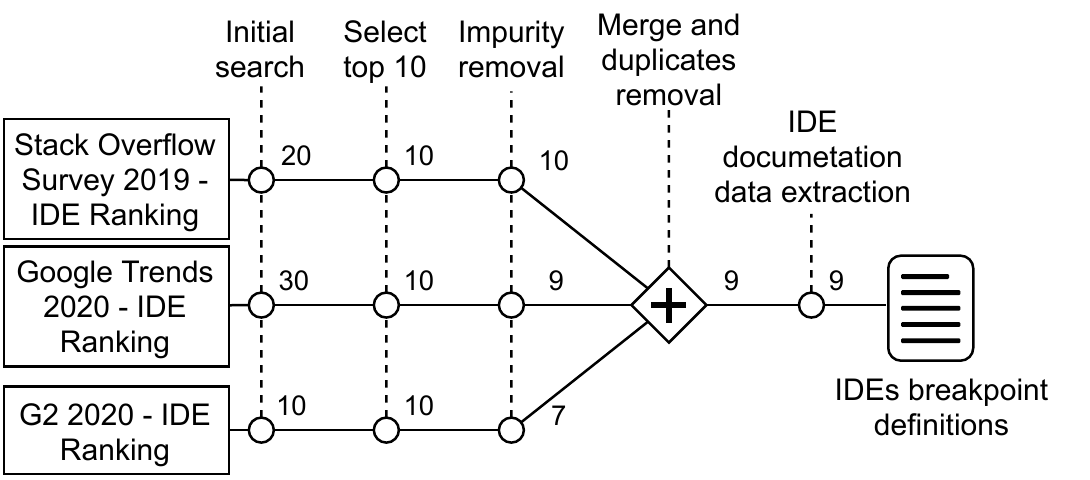}
\caption{\textbf{IDEs selection process and definition extraction} -- We selected nine IDEs by crossing the IDE rankings according to the Stack Overflow Survey 2019, Google Trends 2020, and G2 2020 websites. From the selected IDEs, we extracted breakpoint type definitions in IDEs documentation.}
\label{fig:figIdeSelectionProcess}
\end{figure}

We searched breakpoint type descriptions in IDEs documentation. For that, we indexed the main IDEs considering the crossing of the IDE rankings according to the Stack Overflow Survey 2019 \cite{stackOverflowSurvey2019}, Google Trends 2020 \cite{googleTrends2020}, and G2 2020 \cite{g22020} websites, as presented in Figure \ref{fig:figIdeSelectionProcess}. From the IDEs documentation, we extracted the breakpoint type definitions. We describe below the five steps performed in this process.

\textit{Initial search:} first, we selected the IDEs from the rankings of three websites. The Stack Overflow platform offers a ranking of IDEs that was created through surveys made directly to developers, conducted in 2019. 
We extracted Google Trends 2020 from the Pypl website, which uses the search term metrics provided by Google Trends to generate a ranking of the most searched IDEs until November 2020. We selected the G2 website because it presents a ranking of the most widely used IDEs in the market according to the analysis of technical experts from the G2 itself.

\textit{Select top 10:} we select the top ten IDE from the three rankings because they are widely used or they have easily accessible documentation.

\textit{Impurity removal:} we removed some development tools that are text editors with debugging plugins.

\textit{Merge and duplicates removal:} we combined the IDEs in each ranking by removing duplicates and obtained a new set of nine IDEs.

\textit{IDE documentation data extraction:} finally, we extracted the descriptions of the breakpoint functionalities from the IDE website.

\subsubsection{Step 2 - Academic Literature Data Extraction - RQ2}
\label{subsubsec:Step2AcademicLiteratureDataExtractionRQ2}

\begin{figure}[ht]
\centering
\includegraphics[width=\linewidth]{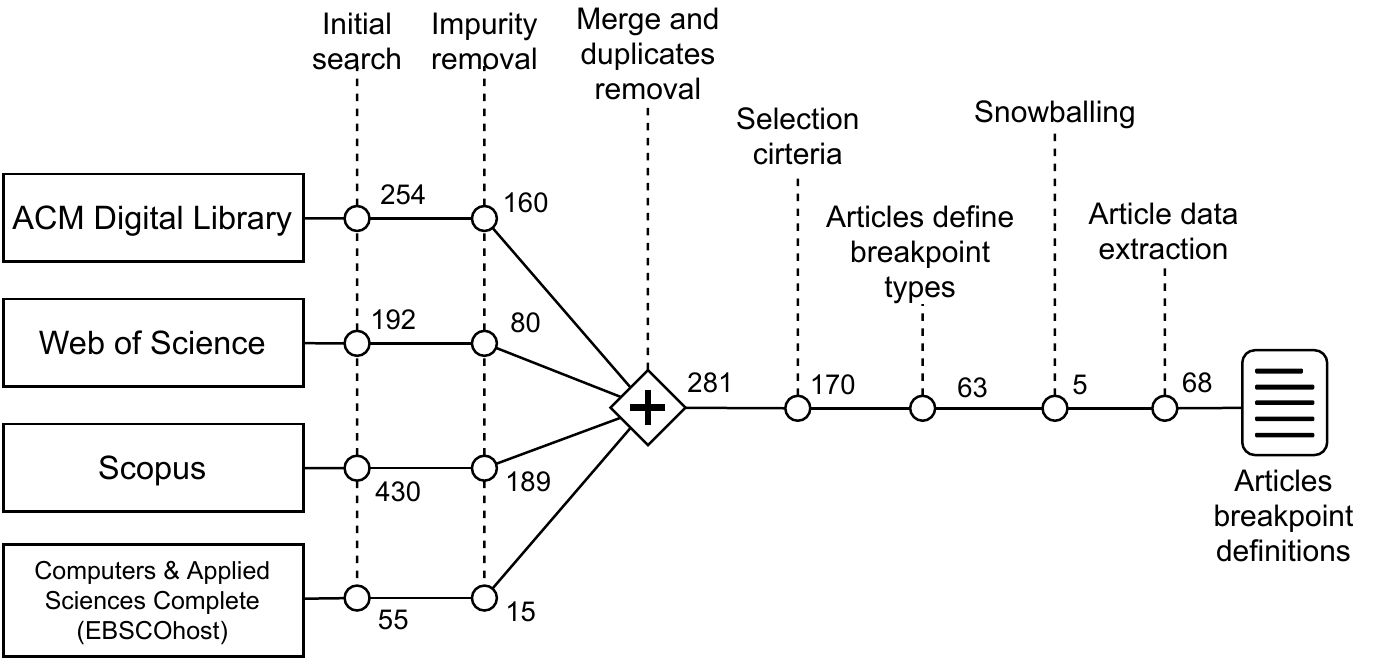}
\caption{\textbf{Articles selection process and definition extraction} -- We selected 68 articles of journals and conference proceedings from specialized academic libraries in software engineering. From the selected articles, we analyzed and extracted breakpoint type definitions.}
\label{fig:figArticleSelectionProcess}
\end{figure}

We indexed 68 articles from journals and conference proceedings that partially classify or partially define breakpoint types, as presented in Figure \ref{fig:figArticleSelectionProcess}. 
In addition, we performed the snowballing search method. Finally, from the selected articles, we extracted the breakpoint type definitions. We describe below the seven steps performed in this process.

\textit{Initial search:} first, we selected the four main scientific research specialized academic libraries in software engineering: ACM Digital Library, Web of Science, Scopus, The Collection of Computer Science Bibliographies and Computers \& Applied Sciences Complete (EBSCOhost). The search query used is presented below. 
The search request was applied to the libraries considering title, abstract, and keywords. In this initial search, we had a total of 931 articles found.

\begin{verbatim}
(breakpoint* OR break-point* 
    OR watchpoint* OR watch-point* 
    OR tracepoint* OR trace-point* 
    OR logpoint* OR log-point* 
    OR catchpoint* OR catch-point*) 
AND debug*
\end{verbatim}

\textit{Impurity removal:} we removed articles that are related to private software tools or articles that describe breakpoints in the context of hardware component debugging. After impurity removal, we had a total of 444 articles.

\textit{Merge and duplicates removal:} we combined all studies into a single dataset. Duplicated entries have been matched by title, authors, year, and type of publication. After applying merge and removal duplicates, we had a total of 281 articles.

\textit{Selection criteria:} over the 281 articles, we selected only journals and conference articles that are full paper with eight pages or more. After applying these selection criteria, we had a total of 170 articles.

\textit{Articles define breakpoint types:} we performed a systematic search over the entire paper texts by the key-words: breakpoint, watchpoint, tracepoints, logpoint, and catchpoint. With this search, we selected the articles with breakpoint type description, partially or completely. After this systematic search, we had a total of 63 articles.

\textit{Snowballing:} from 63 selected articles, we complemented the systematic search with a snowballing search. With snowballing, we added five articles that have descriptions of breakpoint types. In total, we selected 68 articles from which we performed data extraction.

\textit{Article data extraction:} finally, we analyzed the articles and extracted breakpoint type definitions, \ie the breakpoint descriptions, breakpoint type descriptions, and breakpoint functionality descriptions. The extracted data were stored in a spreadsheet. We collected a total of 243 text extracts from 68 articles, which we analyzed together with the text extracts from the IDEs documentation and from the questionnaire data in subsection \ref{subsubsec:Step4AnalysusAndMappingOfDefinitions}.

\subsubsection{Step 3 - Questionnaire Data Extraction - RQ3}
\label{subsubsec:Step3QuestionnaireDataExtractionRQ3}

We administered a questionnaire for breakpoint practitioners to gather opinions and understanding of the breakpoint types used by software developers in the IDEs. 

\textit{Questions:} the questionnaire has 30 questions divided into three sections: 1) general information about the respondent; 2) discursive questions about the developers understanding of the different breakpoint types; and 3) objective questions about understanding the definitions of the different breakpoint types used in the IDE documentation. 

\textit{Participants:} the participants of our questionnaire were included by the following criteria: 1) adult people who have experience using IDE; 2) and adult people who have experience using breakpoints. The participants that did not meet the following criteria were not included: 1) anyone who had no experience using IDE; 2) anyone who had no experience using breakpoints; 3) and anyone who refused to answer the questionnaire.

\textit{Ethics:} the questionnaire was evaluated and consented on February 5, 2021 by the Research Ethics Committee of the University of Quebec at Chicoutimi (CER-UQAC) and certificated by the Ethics Certificate Reference No: 2021-690.

\textit{Dissemination:} we disseminated the questionnaire through social media, software development communities on Discord servers, and directly to software developers. All responses were anonymous. 

\textit{Responses:} we got a total of 70 responses, 29 of which were valid, \ie form completed by the end, and 41 were invalid, \ie partially answered form. Only the 29 valid responses were considered for analysis in subsection \ref{subsubsec:Step4AnalysusAndMappingOfDefinitions}.

\subsubsection{Step 4 - Analysis and Mapping of Definitions}
\label{subsubsec:Step4AnalysusAndMappingOfDefinitions}

In the data analysis subprocess, we applied The Grounded Theory (GT) method. The Grounded Theory is an interpretive and qualitative research method that uses a systematic set of procedures to develop an inductively derived theory about a phenomenon \cite{glaser2017discovery, charmaz2006constructing}. 
The set of procedures are composed by three steps: Open coding, Axial coding, and Selective coding. The steps are complementary to each other and they are executed in an iterative way. Details of the result of each Grounded Theory step are described in section \ref{sec:breakpointMapping}.

In the \textit{Open coding} step, we analyzed definitions, features, properties, and characteristics of the breakpoint types described in the bibliography. We defined attributes, categories, and subcategories, and we use them to classify the definitions of the concepts found. In the \textit{Axial coding} step, we mapped each breakpoint type description classifying by the attributes. We organized the attributes hierarchically to form a logical structure of relationships and similarity among the extracted concepts. In the \textit{Selective coding} step, based on the feature similarities and relationships, we refined the categorization, and we formulated theories to characterize each of the categories.

\section{Breakpoint Mapping}
\label{sec:breakpointMapping}

In this section, we present the artifacts and results from The Grounded Theory data analysis and mapping performed. 

\begin{table*}[p]
    \caption{Breakpoint Types on IDEs}
    \centering
    \setlength{\tabcolsep}{0.3em} 
    {\renewcommand{\arraystretch}{1.3} 
        \begin{tabular}{|llllllllllc|}
        \hline
        \rowcolor[HTML]{BDBDBD} 
        \textbf{Breakpoint Terms}        & \textbf{NetBeans}   & \textbf{Eclipse}    & \textbf{Visual Studio} & \textbf{Visual Studio Code}      & \textbf{IntelliJ}   & \textbf{PhpStorm}   & \textbf{Xcode}      & \textbf{PyCharm}    & \textbf{Android Studio} & \multicolumn{1}{l|}{\cellcolor[HTML]{BDBDBD}\textbf{Total}} \\\hline
        \rowcolor[HTML]{FFFFFF} 
        Breakpoint   & \multicolumn{1}{c}{\cellcolor[HTML]{FFFFFF}\checkmark} & \multicolumn{1}{c}{\cellcolor[HTML]{FFFFFF}\checkmark} & \multicolumn{1}{c}{\cellcolor[HTML]{FFFFFF}\checkmark} & \multicolumn{1}{c}{\cellcolor[HTML]{FFFFFF}\checkmark} & \multicolumn{1}{c}{\cellcolor[HTML]{FFFFFF}\checkmark} & \multicolumn{1}{c}{\cellcolor[HTML]{FFFFFF}\checkmark} & \multicolumn{1}{c}{\cellcolor[HTML]{FFFFFF}\checkmark} & \multicolumn{1}{c}{\cellcolor[HTML]{FFFFFF}\checkmark} & \multicolumn{1}{c}{\cellcolor[HTML]{FFFFFF}\checkmark} & 9  \\
        \rowcolor[HTML]{F3F3F3} 
        Exception Breakpoint   & \multicolumn{1}{c}{\cellcolor[HTML]{F3F3F3}\checkmark} & \multicolumn{1}{c}{\cellcolor[HTML]{F3F3F3}\checkmark} & & & \multicolumn{1}{c}{\cellcolor[HTML]{F3F3F3}\checkmark} & \multicolumn{1}{c}{\cellcolor[HTML]{F3F3F3}\checkmark} & \multicolumn{1}{c}{\cellcolor[HTML]{F3F3F3}\checkmark} & \multicolumn{1}{c}{\cellcolor[HTML]{F3F3F3}\checkmark} & \multicolumn{1}{c}{\cellcolor[HTML]{F3F3F3}\checkmark} & 7  \\
        \rowcolor[HTML]{FFFFFF} 
        Line Breakpoint     & \multicolumn{1}{c}{\cellcolor[HTML]{FFFFFF}\checkmark} & \multicolumn{1}{c}{\cellcolor[HTML]{FFFFFF}\checkmark} & & & \multicolumn{1}{c}{\cellcolor[HTML]{FFFFFF}\checkmark} & \multicolumn{1}{c}{\cellcolor[HTML]{FFFFFF}\checkmark} & & \multicolumn{1}{c}{\cellcolor[HTML]{FFFFFF}\checkmark} & & 5  \\
        \rowcolor[HTML]{F3F3F3} 
        Method Breakpoint   & \multicolumn{1}{c}{\cellcolor[HTML]{F3F3F3}\checkmark} & \multicolumn{1}{c}{\cellcolor[HTML]{F3F3F3}\checkmark} & & & \multicolumn{1}{c}{\cellcolor[HTML]{F3F3F3}\checkmark} & \multicolumn{1}{c}{\cellcolor[HTML]{F3F3F3}\checkmark} & \multicolumn{1}{c}{\cellcolor[HTML]{F3F3F3}\checkmark} & & & 5  \\
        \rowcolor[HTML]{FFFFFF} 
        Tracepoint   & & \multicolumn{1}{c}{\cellcolor[HTML]{FFFFFF}\checkmark} & \multicolumn{1}{c}{\cellcolor[HTML]{FFFFFF}\checkmark} & & & & & & & 2  \\
        \rowcolor[HTML]{F3F3F3} 
        Object Breakpoint   & & & \multicolumn{1}{c}{\cellcolor[HTML]{F3F3F3}\checkmark} & & \multicolumn{1}{c}{\cellcolor[HTML]{F3F3F3}\checkmark} & & & & & 2  \\
        \rowcolor[HTML]{FFFFFF} 
        Function Breakpoint & & & \multicolumn{1}{c}{\cellcolor[HTML]{FFFFFF}\checkmark} & \multicolumn{1}{c}{\cellcolor[HTML]{FFFFFF}\checkmark} & & & & & & 2  \\
        \rowcolor[HTML]{F3F3F3} 
        Data Breakpoint     & & & \multicolumn{1}{c}{\cellcolor[HTML]{F3F3F3}\checkmark} & \multicolumn{1}{c}{\cellcolor[HTML]{F3F3F3}\checkmark} & & & & & & 2  \\
        \rowcolor[HTML]{FFFFFF} 
        Breakpoint Log      & & & & & & \multicolumn{1}{c}{\cellcolor[HTML]{FFFFFF}\checkmark} & & \multicolumn{1}{c}{\cellcolor[HTML]{FFFFFF}\checkmark} & & 2  \\
        \rowcolor[HTML]{F3F3F3} 
        Breakpoint Stack Trace & & & & & & \multicolumn{1}{c}{\cellcolor[HTML]{F3F3F3}\checkmark} & & \multicolumn{1}{c}{\cellcolor[HTML]{F3F3F3}\checkmark} & & 2  \\
        \rowcolor[HTML]{FFFFFF} 
        Class Breakpoint    & \multicolumn{1}{c}{\cellcolor[HTML]{FFFFFF}\checkmark} & \multicolumn{1}{c}{\cellcolor[HTML]{FFFFFF}\checkmark} & & & & & & & & 2  \\
        \rowcolor[HTML]{F3F3F3} 
        Watchpoint   & & \multicolumn{1}{c}{\cellcolor[HTML]{F3F3F3}\checkmark} & & & & & & & \multicolumn{1}{c}{\cellcolor[HTML]{F3F3F3}\checkmark} & 2  \\
        \rowcolor[HTML]{FFFFFF} 
        Data Breakpoint C++ & & & \multicolumn{1}{c}{\cellcolor[HTML]{FFFFFF}\checkmark} & & & & & & & 1  \\
        \rowcolor[HTML]{F3F3F3} 
        Logpoint     & & & & \multicolumn{1}{c}{\cellcolor[HTML]{F3F3F3}\checkmark} & & & & & & 1  \\
        \rowcolor[HTML]{FFFFFF} 
        Conditional Breakpoint & & & & \multicolumn{1}{c}{\cellcolor[HTML]{FFFFFF}\checkmark} & & & & & & 1  \\
        \rowcolor[HTML]{F3F3F3} 
        Inline Breakpoint   & & & & \multicolumn{1}{c}{\cellcolor[HTML]{F3F3F3}\checkmark} & & & & & & 1  \\
        \rowcolor[HTML]{FFFFFF} 
        Field Watchpoint    & & & & & \multicolumn{1}{c}{\cellcolor[HTML]{FFFFFF}\checkmark} & & & & & 1  \\
        \rowcolor[HTML]{F3F3F3} 
        Field Breakpoint    & \multicolumn{1}{c}{\cellcolor[HTML]{F3F3F3}\checkmark} & & & & & & & & & 1  \\
        \rowcolor[HTML]{FFFFFF} 
        Variable Breakpoint & \multicolumn{1}{c}{\cellcolor[HTML]{FFFFFF}\checkmark} & & & & & & & & & 1  \\
        \rowcolor[HTML]{F3F3F3} 
        Thread Breakpoint   & \multicolumn{1}{c}{\cellcolor[HTML]{F3F3F3}\checkmark} & & & & & & & & & 1  \\
        \rowcolor[HTML]{FFFFFF} 
        Logging Breakpoint  & \multicolumn{1}{c}{\cellcolor[HTML]{FFFFFF}\checkmark} & & & & & & & & & 1  \\
        \rowcolor[HTML]{F3F3F3} 
        Event Breakpoint    & & & & & & & \multicolumn{1}{c}{\cellcolor[HTML]{F3F3F3}\checkmark} & & & 1  \\
        \rowcolor[HTML]{FFFFFF} 
        Symbolic Breakpoint & & & & & & & \multicolumn{1}{c}{\cellcolor[HTML]{FFFFFF}\checkmark} & & & 1  \\
        \rowcolor[HTML]{F3F3F3} 
        Test Failure Breakpoint & & & & & & & \multicolumn{1}{c}{\cellcolor[HTML]{F3F3F3}\checkmark} & & & 1  \\\hline
        \rowcolor[HTML]{FFFFFF} 
        \multicolumn{1}{|r}{\cellcolor[HTML]{FFFFFF}\textbf{Total}} & \multicolumn{1}{c}{\cellcolor[HTML]{FFFFFF}9} & \multicolumn{1}{c}{\cellcolor[HTML]{FFFFFF}7} & \multicolumn{1}{c}{\cellcolor[HTML]{FFFFFF}6} & \multicolumn{1}{c}{\cellcolor[HTML]{FFFFFF}6} & \multicolumn{1}{c}{\cellcolor[HTML]{FFFFFF}6} & \multicolumn{1}{c}{\cellcolor[HTML]{FFFFFF}6} & \multicolumn{1}{c}{\cellcolor[HTML]{FFFFFF}6} & \multicolumn{1}{c}{\cellcolor[HTML]{FFFFFF}5} & \multicolumn{1}{c}{\cellcolor[HTML]{FFFFFF}3} & \multicolumn{1}{l|}{\cellcolor[HTML]{FFFFFF}} 
        \\\hline
        \end{tabular}
    }
    \label{tab:tabBreakpointTermsInIDEs}
\end{table*}

\begin{table*}[p]
    \fontsize{7}{8.4}\selectfont
    \caption{Breakpoint Terms on Articles}
    \centering
    \setlength{\tabcolsep}{0.24em} 
    {\renewcommand{\arraystretch}{1.3} 
        \begin{tabular}{|l!{\color[gray]{0.5}\vrule width 0.2pt}l!{\color[gray]{0.5}\vrule width 0.2pt}l!{\color[gray]{0.5}\vrule width 0.2pt}l!{\color[gray]{0.5}\vrule width 0.2pt}l!{\color[gray]{0.5}\vrule width 0.2pt}l!{\color[gray]{0.5}\vrule width 0.2pt}l!{\color[gray]{0.5}\vrule width 0.2pt}l!{\color[gray]{0.5}\vrule width 0.2pt}l!{\color[gray]{0.5}\vrule width 0.2pt}l!{\color[gray]{0.5}\vrule width 0.2pt}l!{\color[gray]{0.5}\vrule width 0.2pt}l!{\color[gray]{0.5}\vrule width 0.2pt}l!{\color[gray]{0.5}\vrule width 0.2pt}l!{\color[gray]{0.5}\vrule width 0.2pt}l!{\color[gray]{0.5}\vrule width 0.2pt}l!{\color[gray]{0.5}\vrule width 0.2pt}l!{\color[gray]{0.5}\vrule width 0.2pt}l!{\color[gray]{0.5}\vrule width 0.2pt}l!{\color[gray]{0.5}\vrule width 0.2pt}l!{\color[gray]{0.5}\vrule width 0.2pt}l!{\color[gray]{0.5}\vrule width 0.2pt}l!{\color[gray]{0.5}\vrule width 0.2pt}l!{\color[gray]{0.5}\vrule width 0.2pt}l!{\color[gray]{0.5}\vrule width 0.2pt}l!{\color[gray]{0.5}\vrule width 0.2pt}l!{\color[gray]{0.5}\vrule width 0.2pt}l!{\color[gray]{0.5}\vrule width 0.2pt}l!{\color[gray]{0.5}\vrule width 0.2pt}l!{\color[gray]{0.5}\vrule width 0.2pt}l!{\color[gray]{0.5}\vrule width 0.2pt}l!{\color[gray]{0.5}\vrule width 0.2pt}l!{\color[gray]{0.5}\vrule width 0.2pt}l!{\color[gray]{0.5}\vrule width 0.2pt}l!{\color[gray]{0.5}\vrule width 0.2pt}l!{\color[gray]{0.5}\vrule width 0.2pt}l!{\color[gray]{0.5}\vrule width 0.2pt}l!{\color[gray]{0.5}\vrule width 0.2pt}l|}
        \hline
        \rowcolor[HTML]{BDBDBD}
  \textbf{Article} &
  \multicolumn{1}{c!{\color[gray]{0.5}\vrule width 0.2pt}}{\textbf{Year}} &
  \multicolumn{1}{c!{\color[gray]{0.5}\vrule width 0.2pt}}{\rotatebox[origin=c]{90}{Breakpoint}} &
  \multicolumn{1}{c!{\color[gray]{0.5}\vrule width 0.2pt}}{\rotatebox[origin=c]{90}{Conditional}} &
  \multicolumn{1}{c!{\color[gray]{0.5}\vrule width 0.2pt}}{\rotatebox[origin=c]{90}{Watchpoint}} &
  \multicolumn{1}{c!{\color[gray]{0.5}\vrule width 0.2pt}}{\rotatebox[origin=c]{90}{Data}} &
  \multicolumn{1}{c!{\color[gray]{0.5}\vrule width 0.2pt}}{\rotatebox[origin=c]{90}{Dynamic}} &
  \multicolumn{1}{c!{\color[gray]{0.5}\vrule width 0.2pt}}{\rotatebox[origin=c]{90}{Code}} &
  \multicolumn{1}{c!{\color[gray]{0.5}\vrule width 0.2pt}}{\rotatebox[origin=c]{90}{Static}} &
  \multicolumn{1}{c!{\color[gray]{0.5}\vrule width 0.2pt}}{\rotatebox[origin=c]{90}{Exception}} &
  \multicolumn{1}{c!{\color[gray]{0.5}\vrule width 0.2pt}}{\rotatebox[origin=c]{90}{Hardware}} &
  \multicolumn{1}{c!{\color[gray]{0.5}\vrule width 0.2pt}}{\rotatebox[origin=c]{90}{Software}} &
  \multicolumn{1}{c!{\color[gray]{0.5}\vrule width 0.2pt}}{\rotatebox[origin=c]{90}{Line}} &
  \multicolumn{1}{c!{\color[gray]{0.5}\vrule width 0.2pt}}{\rotatebox[origin=c]{90}{Tracepoint}} &
  \multicolumn{1}{c!{\color[gray]{0.5}\vrule width 0.2pt}}{\rotatebox[origin=c]{90}{Method}} &
  \multicolumn{1}{c!{\color[gray]{0.5}\vrule width 0.2pt}}{\rotatebox[origin=c]{90}{Data Access}} &
  \multicolumn{1}{c!{\color[gray]{0.5}\vrule width 0.2pt}}{\rotatebox[origin=c]{90}{Class}} &
  \multicolumn{1}{c!{\color[gray]{0.5}\vrule width 0.2pt}}{\rotatebox[origin=c]{90}{Field}} &
  \multicolumn{1}{c!{\color[gray]{0.5}\vrule width 0.2pt}}{\rotatebox[origin=c]{90}{\parbox{1.4cm}{Expression Watchpoint}}} &
  \multicolumn{1}{c!{\color[gray]{0.5}\vrule width 0.2pt}}{\rotatebox[origin=c]{90}{\parbox{1.4cm}{Transition Watchpoint}}} &
  \multicolumn{1}{c!{\color[gray]{0.5}\vrule width 0.2pt}}{\rotatebox[origin=c]{90}{Stateful}} &
  \multicolumn{1}{c!{\color[gray]{0.5}\vrule width 0.2pt}}{\rotatebox[origin=c]{90}{Temporal}} &
  \multicolumn{1}{c!{\color[gray]{0.5}\vrule width 0.2pt}}{\rotatebox[origin=c]{90}{Declarative}} &
  \multicolumn{1}{c!{\color[gray]{0.5}\vrule width 0.2pt}}{\rotatebox[origin=c]{90}{Message}} &
  \multicolumn{1}{c!{\color[gray]{0.5}\vrule width 0.2pt}}{\rotatebox[origin=c]{90}{Symbol}} &
  \multicolumn{1}{c!{\color[gray]{0.5}\vrule width 0.2pt}}{\rotatebox[origin=c]{90}{\parbox{1.4cm}{Method Message}}} &
  \multicolumn{1}{c!{\color[gray]{0.5}\vrule width 0.2pt}}{\rotatebox[origin=c]{90}{\parbox{1.4cm}{Message conditional}}} &
  \multicolumn{1}{c!{\color[gray]{0.5}\vrule width 0.2pt}}{\rotatebox[origin=c]{90}{\parbox{1.4cm}{Reverse Watchpoint}}} &
  \multicolumn{1}{c!{\color[gray]{0.5}\vrule width 0.2pt}}{\rotatebox[origin=c]{90}{\parbox{1.4cm}{Reverse Expression Watchpoint}}} &
  \multicolumn{1}{c!{\color[gray]{0.5}\vrule width 0.2pt}}{\rotatebox[origin=c]{90}{\parbox{1.4cm}{Conditional Watchpoint}}} &
  \multicolumn{1}{c!{\color[gray]{0.5}\vrule width 0.2pt}}{\rotatebox[origin=c]{90}{Selective}} &
  \multicolumn{1}{c!{\color[gray]{0.5}\vrule width 0.2pt}}{\rotatebox[origin=c]{90}{Source}} &
  \multicolumn{1}{c!{\color[gray]{0.5}\vrule width 0.2pt}}{\rotatebox[origin=c]{90}{\parbox{1.4cm}{Conditional line}}} &
  \multicolumn{1}{c!{\color[gray]{0.5}\vrule width 0.2pt}}{\rotatebox[origin=c]{90}{\parbox{1.4cm}{Local variable Watchpoint}}} &
  \multicolumn{1}{c!{\color[gray]{0.5}\vrule width 0.2pt}}{\rotatebox[origin=c]{90}{Simulated}} &
  \multicolumn{1}{c!{\color[gray]{0.5}\vrule width 0.2pt}}{\rotatebox[origin=c]{90}{\parbox{1.4cm}{Conditional Data}}} &
  \multicolumn{1}{c!{\color[gray]{0.5}\vrule width 0.2pt}}{\rotatebox[origin=c]{90}{Variable}} &
  \multicolumn{1}{c|}{\textbf{Total}} \\\hline
\rowcolor[HTML]{FFFFFF}
\cite{beller2018dichotomy} & 2018  &   & \multicolumn{1}{c!{\color[gray]{0.5}\vrule width 0.2pt}}{\checkmark} &   &   &   &   &   & \multicolumn{1}{c!{\color[gray]{0.5}\vrule width 0.2pt}}{\checkmark} &   &   & \multicolumn{1}{c!{\color[gray]{0.5}\vrule width 0.2pt}}{\checkmark} & \multicolumn{1}{c!{\color[gray]{0.5}\vrule width 0.2pt}}{\checkmark} & \multicolumn{1}{c!{\color[gray]{0.5}\vrule width 0.2pt}}{\checkmark} & \multicolumn{1}{c!{\color[gray]{0.5}\vrule width 0.2pt}}{\checkmark} & \multicolumn{1}{c!{\color[gray]{0.5}\vrule width 0.2pt}}{\checkmark} & \multicolumn{1}{c!{\color[gray]{0.5}\vrule width 0.2pt}}{\checkmark} &   &   &   &   &   &   &   &   &   &   &   &   &   &   &   &   &   &   &   & \multicolumn{1}{c|}{8} \\
\rowcolor[HTML]{F3F3F3}
\cite{dupriez2017analysis} & 2017  & \multicolumn{1}{c!{\color[gray]{0.5}\vrule width 0.2pt}}{\checkmark} & \multicolumn{1}{c!{\color[gray]{0.5}\vrule width 0.2pt}}{\checkmark} & \multicolumn{1}{c!{\color[gray]{0.5}\vrule width 0.2pt}}{\checkmark} &   &   &   &   &   &   &   &   &   &   &   &   &   & \multicolumn{1}{c!{\color[gray]{0.5}\vrule width 0.2pt}}{\checkmark} & \multicolumn{1}{c!{\color[gray]{0.5}\vrule width 0.2pt}}{\checkmark} &   &   &   &   &   &   &   &   &   &   &   &   &   &   &   &   &   & \multicolumn{1}{c|}{5} \\
\rowcolor[HTML]{FFFFFF}
\cite{corrodi2016towards} & 2016  &   & \multicolumn{1}{c!{\color[gray]{0.5}\vrule width 0.2pt}}{\checkmark} &   &   & \multicolumn{1}{c!{\color[gray]{0.5}\vrule width 0.2pt}}{\checkmark} &   &   &   &   &   &   &   &   &   &   &   &   &   & \multicolumn{1}{c!{\color[gray]{0.5}\vrule width 0.2pt}}{\checkmark} & \multicolumn{1}{c!{\color[gray]{0.5}\vrule width 0.2pt}}{\checkmark} & \multicolumn{1}{c!{\color[gray]{0.5}\vrule width 0.2pt}}{\checkmark} &   &   &   &   &   &   &   &   &   &   &   &   &   &   & \multicolumn{1}{c|}{5} \\
\rowcolor[HTML]{F3F3F3}
\cite{johnson1982software} & 1982 &   & \multicolumn{1}{c!{\color[gray]{0.5}\vrule width 0.2pt}}{\checkmark} &   & \multicolumn{1}{c!{\color[gray]{0.5}\vrule width 0.2pt}}{\checkmark} & \multicolumn{1}{c!{\color[gray]{0.5}\vrule width 0.2pt}}{\checkmark} & \multicolumn{1}{c!{\color[gray]{0.5}\vrule width 0.2pt}}{\checkmark} & \multicolumn{1}{c!{\color[gray]{0.5}\vrule width 0.2pt}}{\checkmark} &   &   &   &   &   &   &   &   &   &   &   &   &   &   &   &   &   &   &   &   &   &   &   &   &   &   &   &   & \multicolumn{1}{c|}{5} \\
\rowcolor[HTML]{FFFFFF}
\cite{petrillodevelopers2017} & 2017 & \multicolumn{1}{c!{\color[gray]{0.5}\vrule width 0.2pt}}{\checkmark} & \multicolumn{1}{c!{\color[gray]{0.5}\vrule width 0.2pt}}{\checkmark} & \multicolumn{1}{c!{\color[gray]{0.5}\vrule width 0.2pt}}{\checkmark} &   & \multicolumn{1}{c!{\color[gray]{0.5}\vrule width 0.2pt}}{\checkmark} &   & \multicolumn{1}{c!{\color[gray]{0.5}\vrule width 0.2pt}}{\checkmark} &   &   &   &   &   &   &   &   &   &   &   &   &   &   &   &   &   &   &   &   &   &   &   &   &   &   &   &   & \multicolumn{1}{c|}{5} \\
\rowcolor[HTML]{F3F3F3}
\cite{boix2012handling} & 2012 & \multicolumn{1}{c!{\color[gray]{0.5}\vrule width 0.2pt}}{\checkmark} &   &   &   &   &   &   &   &   &   &   &   &   &   &   &   &   &   &   &   &   & \multicolumn{1}{c!{\color[gray]{0.5}\vrule width 0.2pt}}{\checkmark} & \multicolumn{1}{c!{\color[gray]{0.5}\vrule width 0.2pt}}{\checkmark} & \multicolumn{1}{c!{\color[gray]{0.5}\vrule width 0.2pt}}{\checkmark} & \multicolumn{1}{c!{\color[gray]{0.5}\vrule width 0.2pt}}{\checkmark} &   &   &   &   &   &   &   &   &   &   & \multicolumn{1}{c|}{5} \\
\rowcolor[HTML]{FFFFFF}
\cite{lencevicius1999query} & 1999 & \multicolumn{1}{c!{\color[gray]{0.5}\vrule width 0.2pt}}{\checkmark} & \multicolumn{1}{c!{\color[gray]{0.5}\vrule width 0.2pt}}{\checkmark} &   & \multicolumn{1}{c!{\color[gray]{0.5}\vrule width 0.2pt}}{\checkmark} &   &   &   &   &   &   &   &   &   &   &   &   &   &   &   &   &   &   &   &   &   &   &   &   &   &   &   &   &   & \multicolumn{1}{c!{\color[gray]{0.5}\vrule width 0.2pt}}{\checkmark} &   & \multicolumn{1}{c|}{4} \\
\rowcolor[HTML]{F3F3F3}
\cite{maruyama2003debugging} & 2003 &   & \multicolumn{1}{c!{\color[gray]{0.5}\vrule width 0.2pt}}{\checkmark} & \multicolumn{1}{c!{\color[gray]{0.5}\vrule width 0.2pt}}{\checkmark} &   & \multicolumn{1}{c!{\color[gray]{0.5}\vrule width 0.2pt}}{\checkmark} &   &   &   &   &   &   &   &   &   &   &   &   &   &   &   &   &   &   &   &   & \multicolumn{1}{c!{\color[gray]{0.5}\vrule width 0.2pt}}{\checkmark} &   &   &   &   &   &   &   &   &   & \multicolumn{1}{c|}{4} \\
\rowcolor[HTML]{FFFFFF}
\cite{visan2012temporal} & 2012 & \multicolumn{1}{c!{\color[gray]{0.5}\vrule width 0.2pt}}{\checkmark} &   &   &   &   &   &   &   & \multicolumn{1}{c!{\color[gray]{0.5}\vrule width 0.2pt}}{\checkmark} & \multicolumn{1}{c!{\color[gray]{0.5}\vrule width 0.2pt}}{\checkmark} &   &   &   &   &   &   &   &   &   &   &   &   &   &   &   &   & \multicolumn{1}{c!{\color[gray]{0.5}\vrule width 0.2pt}}{\checkmark} &   &   &   &   &   &   &   &   & \multicolumn{1}{c|}{4} \\
\rowcolor[HTML]{F3F3F3}
\cite{vasudevan2009re} & 2009 &   &   &   & \multicolumn{1}{c!{\color[gray]{0.5}\vrule width 0.2pt}}{\checkmark} &   & \multicolumn{1}{c!{\color[gray]{0.5}\vrule width 0.2pt}}{\checkmark} &   &   & \multicolumn{1}{c!{\color[gray]{0.5}\vrule width 0.2pt}}{\checkmark} & \multicolumn{1}{c!{\color[gray]{0.5}\vrule width 0.2pt}}{\checkmark} &   &   &   &   &   &   &   &   &   &   &   &   &   &   &   &   &   &   &   &   &   &   &   &   &   & \multicolumn{1}{c|}{4} \\
\rowcolor[HTML]{FFFFFF}
\cite{petrillo2019swarm} & 2019  & \multicolumn{1}{c!{\color[gray]{0.5}\vrule width 0.2pt}}{\checkmark} &   & \multicolumn{1}{c!{\color[gray]{0.5}\vrule width 0.2pt}}{\checkmark} &   & \multicolumn{1}{c!{\color[gray]{0.5}\vrule width 0.2pt}}{\checkmark} &   & \multicolumn{1}{c!{\color[gray]{0.5}\vrule width 0.2pt}}{\checkmark} &   &   &   &   &   &   &   &   &   &   &   &   &   &   &   &   &   &   &   &   &   &   &   &   &   &   &   &   & \multicolumn{1}{c|}{4} \\
\rowcolor[HTML]{F3F3F3}
\cite{mcdowell1989debugging} & 1989 & \multicolumn{1}{c!{\color[gray]{0.5}\vrule width 0.2pt}}{\checkmark} & \multicolumn{1}{c!{\color[gray]{0.5}\vrule width 0.2pt}}{\checkmark} &   &   &   &   &   & \multicolumn{1}{c!{\color[gray]{0.5}\vrule width 0.2pt}}{\checkmark} &   &   &   &   &   &   &   &   &   &   &   &   &   &   &   &   &   &   &   &   &   &   &   &   &   &   & \multicolumn{1}{c!{\color[gray]{0.5}\vrule width 0.2pt}}{\checkmark} & \multicolumn{1}{c|}{4} \\
\rowcolor[HTML]{FFFFFF}
\cite{copperman1995poor} & 1995  &   &   & \multicolumn{1}{c!{\color[gray]{0.5}\vrule width 0.2pt}}{\checkmark} & \multicolumn{1}{c!{\color[gray]{0.5}\vrule width 0.2pt}}{\checkmark} &   &   &   &   &   &   &   &   &   &   &   &   &   &   &   &   &   &   &   &   &   &   &   & \multicolumn{1}{c!{\color[gray]{0.5}\vrule width 0.2pt}}{\checkmark} &   &   &   &   &   &   &   & \multicolumn{1}{c|}{3} \\
\rowcolor[HTML]{F3F3F3}
\cite{jahne1996peard} & 1996 & \multicolumn{1}{c!{\color[gray]{0.5}\vrule width 0.2pt}}{\checkmark} & \multicolumn{1}{c!{\color[gray]{0.5}\vrule width 0.2pt}}{\checkmark} &   &   &   &   &   &   &   &   &   &   &   &   &   &   &   &   &   &   &   &   &   &   &   &   &   &   & \multicolumn{1}{c!{\color[gray]{0.5}\vrule width 0.2pt}}{\checkmark} &   &   &   &   &   &   & \multicolumn{1}{c|}{3} \\
\rowcolor[HTML]{FFFFFF}
\cite{kosar2014debugging} & 2014 & \multicolumn{1}{c!{\color[gray]{0.5}\vrule width 0.2pt}}{\checkmark} &   &   & \multicolumn{1}{c!{\color[gray]{0.5}\vrule width 0.2pt}}{\checkmark} &   &   &   &   &   &   &   &   &   &   &   &   &   &   &   &   &   &   &   &   &   &   &   &   &   & \multicolumn{1}{c!{\color[gray]{0.5}\vrule width 0.2pt}}{\checkmark} &   &   &   &   &   & \multicolumn{1}{c|}{3} \\
\rowcolor[HTML]{F3F3F3}
\cite{spinellis2006debuggers} & 2006 &   &   & \multicolumn{1}{c!{\color[gray]{0.5}\vrule width 0.2pt}}{\checkmark} & \multicolumn{1}{c!{\color[gray]{0.5}\vrule width 0.2pt}}{\checkmark} &   & \multicolumn{1}{c!{\color[gray]{0.5}\vrule width 0.2pt}}{\checkmark} &   &   &   &   &   &   &   &   &   &   &   &   &   &   &   &   &   &   &   &   &   &   &   &   &   &   &   &   &   & \multicolumn{1}{c|}{3} \\
\rowcolor[HTML]{FFFFFF}
\cite{seaton2014debugging} & 2014 &   &   &   &   &   &   &   &   &   &   & \multicolumn{1}{c!{\color[gray]{0.5}\vrule width 0.2pt}}{\checkmark} &   &   &   &   &   &   &   &   &   &   &   &   &   &   &   &   &   &   &   & \multicolumn{1}{c!{\color[gray]{0.5}\vrule width 0.2pt}}{\checkmark} & \multicolumn{1}{c!{\color[gray]{0.5}\vrule width 0.2pt}}{\checkmark} &   &   &   & \multicolumn{1}{c|}{3} \\
\rowcolor[HTML]{F3F3F3}
\cite{alsallakh2012visual} & 2012 & \multicolumn{1}{c!{\color[gray]{0.5}\vrule width 0.2pt}}{\checkmark} & \multicolumn{1}{c!{\color[gray]{0.5}\vrule width 0.2pt}}{\checkmark} &   &   &   &   &   &   &   &   &   & \multicolumn{1}{c!{\color[gray]{0.5}\vrule width 0.2pt}}{\checkmark} &   &   &   &   &   &   &   &   &   &   &   &   &   &   &   &   &   &   &   &   &   &   &   & \multicolumn{1}{c|}{3} \\
\rowcolor[HTML]{FFFFFF}
\cite{kumar2020amber} & 2020 & \multicolumn{1}{c!{\color[gray]{0.5}\vrule width 0.2pt}}{\checkmark} & \multicolumn{1}{c!{\color[gray]{0.5}\vrule width 0.2pt}}{\checkmark} &   &   &   &   &   &   &   &   &   &   &   &   &   &   &   &   &   &   &   &   &   &   &   &   &   &   &   &   &   &   & \multicolumn{1}{c!{\color[gray]{0.5}\vrule width 0.2pt}}{\checkmark} &   &   & \multicolumn{1}{c|}{3} \\\hline
  \multicolumn{1}{|c}{} &
  \multicolumn{1}{c!{\color[gray]{0.5}\vrule width 0.2pt}}{\textbf{Total}} &
  \multicolumn{1}{c!{\color[gray]{0.5}\vrule width 0.2pt}}{11} &
  \multicolumn{1}{c!{\color[gray]{0.5}\vrule width 0.2pt}}{11} &
  \multicolumn{1}{c!{\color[gray]{0.5}\vrule width 0.2pt}}{6} &
  \multicolumn{1}{c!{\color[gray]{0.5}\vrule width 0.2pt}}{6} &
  \multicolumn{1}{c!{\color[gray]{0.5}\vrule width 0.2pt}}{5} &
  \multicolumn{1}{c!{\color[gray]{0.5}\vrule width 0.2pt}}{3} &
  \multicolumn{1}{c!{\color[gray]{0.5}\vrule width 0.2pt}}{3} &
  \multicolumn{1}{c!{\color[gray]{0.5}\vrule width 0.2pt}}{2} &
  \multicolumn{1}{c!{\color[gray]{0.5}\vrule width 0.2pt}}{2} &
  \multicolumn{1}{c!{\color[gray]{0.5}\vrule width 0.2pt}}{2} &
  \multicolumn{1}{c!{\color[gray]{0.5}\vrule width 0.2pt}}{2} &
  \multicolumn{1}{c!{\color[gray]{0.5}\vrule width 0.2pt}}{2} &
  \multicolumn{1}{c!{\color[gray]{0.5}\vrule width 0.2pt}}{1} &
  \multicolumn{1}{c!{\color[gray]{0.5}\vrule width 0.2pt}}{1} &
  \multicolumn{1}{c!{\color[gray]{0.5}\vrule width 0.2pt}}{1} &
  \multicolumn{1}{c!{\color[gray]{0.5}\vrule width 0.2pt}}{1} &
  \multicolumn{1}{c!{\color[gray]{0.5}\vrule width 0.2pt}}{1} &
  \multicolumn{1}{c!{\color[gray]{0.5}\vrule width 0.2pt}}{1} &
  \multicolumn{1}{c!{\color[gray]{0.5}\vrule width 0.2pt}}{1} &
  \multicolumn{1}{c!{\color[gray]{0.5}\vrule width 0.2pt}}{1} &
  \multicolumn{1}{c!{\color[gray]{0.5}\vrule width 0.2pt}}{1} &
  \multicolumn{1}{c!{\color[gray]{0.5}\vrule width 0.2pt}}{1} &
  \multicolumn{1}{c!{\color[gray]{0.5}\vrule width 0.2pt}}{1} &
  \multicolumn{1}{c!{\color[gray]{0.5}\vrule width 0.2pt}}{1} &
  \multicolumn{1}{c!{\color[gray]{0.5}\vrule width 0.2pt}}{1} &
  \multicolumn{1}{c!{\color[gray]{0.5}\vrule width 0.2pt}}{1} &
  \multicolumn{1}{c!{\color[gray]{0.5}\vrule width 0.2pt}}{1} &
  \multicolumn{1}{c!{\color[gray]{0.5}\vrule width 0.2pt}}{1} &
  \multicolumn{1}{c!{\color[gray]{0.5}\vrule width 0.2pt}}{1} &
  \multicolumn{1}{c!{\color[gray]{0.5}\vrule width 0.2pt}}{1} &
  \multicolumn{1}{c!{\color[gray]{0.5}\vrule width 0.2pt}}{1} &
  \multicolumn{1}{c!{\color[gray]{0.5}\vrule width 0.2pt}}{1} &
  \multicolumn{1}{c!{\color[gray]{0.5}\vrule width 0.2pt}}{1} &
  \multicolumn{1}{c!{\color[gray]{0.5}\vrule width 0.2pt}}{1} &
  \multicolumn{1}{c!{\color[gray]{0.5}\vrule width 0.2pt}}{1} &
  \multicolumn{1}{l|}{}
        \\\hline
        \end{tabular}
    }
    \label{tab:tabArticlesBreakpoints}
\end{table*}

\subsection{Open coding output}
\label{subsec:breakpointMappingOpenCoding}
In The Grounded Theory Open coding step, we analyzed the definitions, the features, the properties, and the characteristics of the breakpoint types described both in the IDEs documentation, selected articles, and from the questionnaire. From this, we extracted 39 attributes grouped into four categories, as presented below:

\begin{itemize}
    \item \textbf{Defined at:} line of code, statement in a line of code, by configuration, over object, over property object, field, global variable, local variable, over a message, written in code, block of memory;
    \item \textbf{Triggered by:} line / statement hit, data changes, memory address changes, expression satisfied, expression value change, expression satisfied an object id, conditional hit, hit count, event triggered, method / function called, method enter, method exit, class is loaded, exception throw, field read, field write, test failure, message reaches the head queue, thread start / terminated;
    \item \textbf{Effect:} does not stop execution, stop execution, show log message;
    \item \textbf{Complementary:} conditional as derivative action, has derivative actions, is a derivative action, has hardware support, multithread, not enough information.
\end{itemize}

\begin{figure}[ht]
\centering
\includegraphics[width=\linewidth]{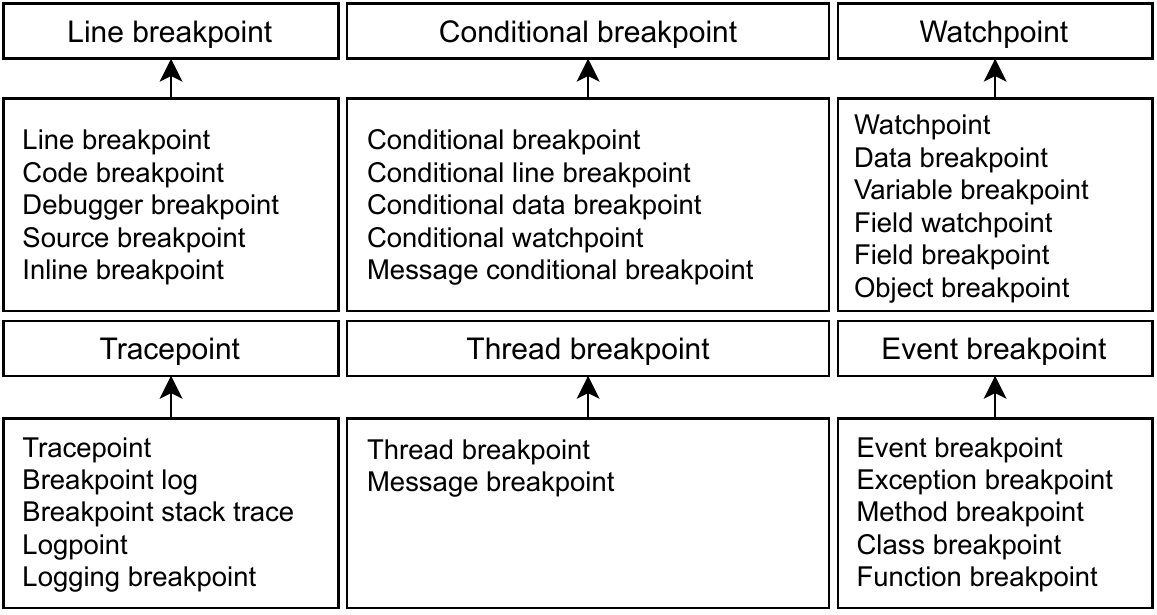}
\caption{\textbf{The Grounded Theory Open Output} -- The main breakpoint types and respective extracted term grouping.}
\label{fig:figTermGrouping}
\end{figure}

\begin{figure}[ht]
\centering
\includegraphics[width=\linewidth]{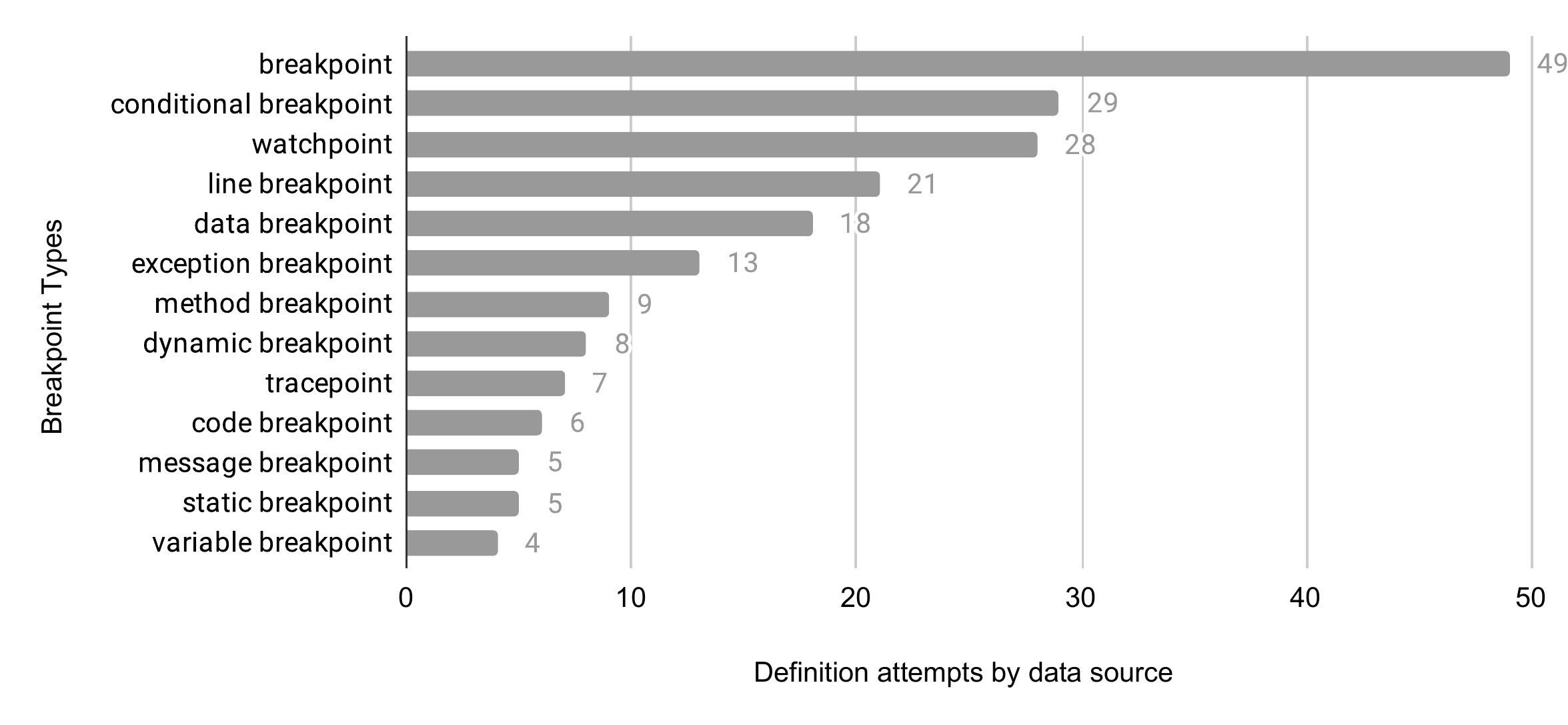}
\caption{\textbf{The Grounded Theory Axial Output} -- Number of times that all data sources presented some description of each breakpoint type.}
\label{fig:figTermOccurrences}
\end{figure}

\begin{table}[ht]
    \caption{The  Grounded  Theory  Axial  Output -- continuation}
    \centering
    \setlength{\tabcolsep}{0.3em} 
    {\renewcommand{\arraystretch}{1.3} 
        \begin{tabular}{|p{1.2cm}p{7.1cm}|}
        \hline
\textbf{Described 3 times} & class, debugger, event, field watchpoint, hardware, simulated, software \\ \hline
\textbf{Described 2 times} & log, stack trace, conditional data, field, function, logpoint, object, source \\ \hline
\textbf{Described 1 time} & behavioral watchpoint, collective, concurrent, conditional line, conditional watchpoint, control, data access, data c++, declarative, declarative tracepoint, expression watchpoint, expression-transition watchpoint, inline, lambda, local variable watchpoint, logging, message conditional, meta-breakpoint, method message, pseudo-breakpoint, reverse expression watchpoint, reverse watchpoint, scan-points, selective, shared, stateful, statement, symbol, symbolic, temporal, test failure, thread, transition watchpoint, unconditional \\ \hline
\multicolumn{2}{l}{$^{\mathrm{a}}$This table is the continuation of the chart in Figure \ref{fig:figTermOccurrences}.}
\\
\multicolumn{2}{l}{$^{\mathrm{b}}$Breakpoint types that have been described three times or less}
        \end{tabular}
    }
    \label{tab:tabTermOccurrences}
\end{table}

The attribute category \textit{Defined At} aggregates the group of attributes that characterize the place or way in which the breakpoint is added in the source code. That is, added in a line of code, through configuration, in a variable or others. The attribute category \textit{Triggered By} aggregates the group of attributes that characterize how the breakpoint is triggered after being added to the source code. That is, triggered by the execution of a tuple of code, by data changing, by triggering an event or other way. The attribute category \textit{Effect} aggregates the group of attributes that characterize the behavior presented by the debugger and by the IDE once the breakpoint is triggered. In other words, if it interrupts the system execution or not or if it presents a log message. The attribute category \textit{Complementary} aggregates the group of attributes that complement the characteristics of each breakpoint and that do not fit into the three previous categories. For example, if the breakpoint type has: conditional, derived actions, hardware support, and others. 

With the attributes defined, we related each attribute to each of the breakpoint type description taken from the documentation, articles, and questionnaire. As a resulting artifact, we obtained a \textit{Mapping} table with the relation of each attribute to each breakpoint type extracted from each of the data source descriptions. See Zenodo repository \cite{andreetta_fontana_eduardo_2021_5213971}.

\subsection{Axial coding output}
\label{subsec:breakpointMappingAxialCoding}
For each of the breakpoint type descriptions, we extracted the terms of the breakpoint types, and we related them with the attributes.

In Figure \ref{fig:figTermOccurrences} and in Table \ref{tab:tabTermOccurrences}, we present a compilation of the breakpoint type terms used by the three data sources. The chart in Figure \ref{fig:figTermOccurrences} shows the number of times that all data sources presented some description to try to define each breakpoint type. For breakpoint types that data sources define three or less times, we group them in Table \ref{tab:tabTermOccurrences}, \ie the table is the chart continuation.

In addition, we created two breakpoint type mapping tables: 1) Table \ref{tab:tabBreakpointTermsInIDEs}: breakpoint types on the IDEs, and 2) Table \ref{tab:tabArticlesBreakpoints}: breakpoint types on articles. For both tables, the total column is the count of incidences that a given term is described by the IDE or article as a breakpoint type. The total line is the count of breakpoint types that each IDE or article describes.

Considering the incidence of descriptions by terms presented in the compilation in Figure \ref{fig:figTermOccurrences} and the \textit{Mapping} table in the dataset in the Zenodo repository \cite{andreetta_fontana_eduardo_2021_5213971}, we identified terms with similar distribution of attributes and high degree of similarity among the characteristics of each term.
From this, we extracted the terms that represent the main breakpoint types: Line Breakpoint, Conditional Breakpoint, Thread Breakpoint, Event Breakpoint, Watchpoint, Tracepoint. 
The Figure \ref{fig:figTermGrouping} shows the term grouping. The box at the top of arrow contains main breakpoint types and the box below the arrow contains the extracted terms with high degree of similarity.

The term Breakpoint is described by the IDEs documentation and the articles in a generic way. Thus, we do not include the term Breakpoint as a breakpoint type. We observed that the term Line Breakpoint is the second one in the Table \ref{tab:tabBreakpointTermsInIDEs} that represents the basic breakpoint type.
The terms Software Breakpoint, Hardware Breakpoint, Static Breakpoint, and Dynamic Breakpoint are not breakpoint types, but rather intermediate categories that suggest the grouping of some breakpoint types.
Complementarily, the term Dynamic Breakpoint groups breakpoint types that depend on events or conditions to be triggered, while the term Static Breakpoint groups breakpoint types that are triggered without dependence on other factors.

The term Conditional Breakpoint is sometimes described as a breakpoint type or at other times as a complementary action of some other breakpoint type. The term Watchpoint is the grouping of all terms that define breakpoint types related to data change, either directly in memory or object instance. The term Tracepoint group all the terms that define breakpoint types that do not stop the system running when triggered, but logging a message. 
The term Event Breakpoint grouped all the terms that define breakpoint types that are triggered when a specific event happens during execution, such as calling a specific method, triggering an exception, instantiating a class or calling a specific function. Thread Breakpoint accurately represents the breakpoint types applied to distributed programming and multithreading, since the term Message Breakpoint is misinterpreted with the breakpoint type that emits log messages, such as Tracepoint.

\subsection{Selective coding output}
\label{subsec:breakpointMappingSelectiveCoding}
The output of Selective coding step are structured definitions for each of the six breakpoint types. We extracted the definitions from relevant terms structuring by the four attribute categories: \textit{Defined At, Triggered By, Effect, and Complementary}.
The definitions are presented in boxes \ref{box:boxLineBreakpoint}, \ref{box:boxConditionalBreakpoint}, \ref{box:boxTracepoint}, \ref{box:boxThreadBreakpoint}, \ref{box:boxEventBreakpoint}, and \ref{box:boxWatchpoint}. We evaluated the definitions from the questionnaire given to software developers. Each respondent gave a score from one to ten for each of the definitions -- being one to totally disagree and ten to totally agree. The arithmetic average of the scores obtained is presented in Table \ref{tab:tabDescriptionScore}.

\subsection{Research questions output}
\label{subsec:breakpointMappingResearchQuestions}
\textbf{RQ1 - What breakpoint types are found in IDEs?}
We extracted all distinct breakpoint type names related to the IDEs analyzed from the \textit{Mapping} table in dataset \cite{andreetta_fontana_eduardo_2021_5213971}. They are presented in Table \ref{tab:tabBreakpointTermsInIDEs}.

\textbf{RQ2 - What breakpoint types are studied in the academic literature?} 
We filtered and extracted all distinct breakpoint type names related to the academic literature from the \textit{Mapping} table in dataset \cite{andreetta_fontana_eduardo_2021_5213971}. The Table \ref{tab:tabArticlesBreakpoints} shows the academic articles that mention three or more breakpoint type names. The complete list is found in the \textit{GT Compilation} spreadsheet in dataset \cite{andreetta_fontana_eduardo_2021_5213971}.

\textbf{RQ3 - What are the breakpoint types according to practitioners?}
We filtered from the all distinct breakpoint type names related to the questionnaire from the \textit{Mapping} table in dataset \cite{andreetta_fontana_eduardo_2021_5213971}. The new terms we found through the questionnaire are:

\begin{itemize}
    \item \textbf{Debugger Breakpoint:} command \textit{debugger;} written in the source code, as the language's own command, to force the system to stop running.
    \item \textbf{Statement Breakpoint:} similar to Lambda or Inline breakpoint. A breakpoint over part of the line of code, or execution tuple.
    \item \textbf{Lambda Breakpoint:} similar to Inline breakpoint. A breakpoint over part of the line of code, or execution tuple.
    \item \textbf{Shared Breakpoint:} similar to Tracepoint. A breakpoint that logs messages and does not stop the system from running. 
\end{itemize}

The sample of questionnaire respondents is composed of 29 different respondents. 
The average years of experience with software development was 9.4 years. From the total sample, 68.96\% reported that, according to their work or university experience, they have studied or have been introduced to breakpoint types.

\begin{definition}[Line Breakpoint]\label{box:boxLineBreakpoint}
\fontsize{9}{10.8}\selectfont
\textbf{Line Breakpoint} is the breakpoint type associated with a line of code.
\textit{Where}: It is usually inserted over the line of code or eventually over a code tuple contained in the line of code, or else written directly in the code as a code stop instruction.
\textit{Trigger}: It is usually triggered when the line of code is hit during the system execution.
\textit{Effect}: When triggered, it causes the system execution interruption.
\end{definition}

\begin{definition}[Conditional Breakpoint]\label{box:boxConditionalBreakpoint}
\fontsize{9}{10.8}\selectfont
\textbf{Conditional Breakpoint} is a breakpoint type associated with an expression.
\textit{Where}: It is usually inserted as a specific breakpoint type over a line of code, or it could be associated with other breakpoint type such as a configuration usually called property or action.
\textit{Trigger}: It is triggered whenever the condition related to its associated expression is true. This condition could be related to data, a count or the existence of a specific object.
\textit{Effect}: When triggered, it causes the system execution interruption.
\end{definition}

\begin{definition}[Thread Breakpoint]\label{box:boxThreadBreakpoint}
\fontsize{9}{10.8}\selectfont
\textbf{Thread Breakpoint} is a breakpoint type applied to distributed systems and parallel programming, associated with threads.
\textit{Where}: It is usually related to a thread or thread message through a configuration option.
\textit{Trigger}: It is usually triggered when the thread with which it is associated starts or stops.
\textit{Effect}: When associated with a thread message, it is triggered when the message with which it is associated is at the top of the execution queue. When triggered, it causes the entire system execution interruption or just the thread with which it is associated.
\textit{Complementary}: Furthermore, it could contain associated expressions that determines its activation only when the condition is true.
\end{definition}

\begin{definition}[Event Breakpoint]\label{box:boxEventBreakpoint}
\fontsize{9}{10.8}\selectfont
\textbf{Event Breakpoint} is a breakpoint type that is system execution event oriented.
\textit{Where}: It is usually inserted by configuration option and eventually over a specific line of code.
\textit{Trigger}: It is usually triggered when the event to which it is associated is triggered by the system, such as calling a method, creating an object of a specific class, or throwing an exception.
\textit{Effect}: When triggered, it causes the system execution interruption.
\textit{Complementary}: Furthermore, it could contain associated expressions that determines its activation only when the condition is true.
\end{definition}

\begin{definition}[Watchpoint]\label{box:boxWatchpoint}
\fontsize{9}{10.8}\selectfont
\textbf{Watchpoint} is a type of data-oriented breakpoint and could be directly related to a memory address.
\textit{Where}: It is usually inserted over a line of code that contains a field or variable or through configuration options.
\textit{Trigger}: It is usually triggered when the data it is associated with is read, written or changed. It could also be triggered when the memory address to which it is associated is read, written or changed.
\textit{Effect}: When triggered, it causes the system execution interruption.
\textit{Complementary}: Furthermore, it could contain associated expressions that determines its activation only when the condition is true.
\end{definition}

\begin{definition}[Tracepoint]\label{box:boxTracepoint}
\fontsize{9}{10.8}\selectfont
\textbf{Tracepoint} logs a message on console output when execution through it. \textit{Where}: It is usually inserted over the line of code.
\textit{Trigger}: It is usually triggered when the line of code with which it is associated is the next line to be executed by the system.
\textit{Effect}: When triggered, it displays a previously defined debug message on the console, \textbf{not interrupting the system execution}.
\textit{Complementary}: Furthermore, it could contain associated expressions that determines its activation only when the condition is true.
\end{definition}

\begin{table}[ht]
    \caption{Breakpoint Type Definition Score}
    \centering
    \setlength{\tabcolsep}{0.2em} 
    {\renewcommand{\arraystretch}{1.2} 
        \begin{tabular}{|lc|}
        \hline
        \rowcolor[HTML]{BDBDBD} 
        \textbf{Definition of} & \multicolumn{1}{c|}{\cellcolor[HTML]{BDBDBD}\textbf{Score}} \\
        \rowcolor[HTML]{FFFFFF} 
        Line Breakpoint & 8.3 \\
        \rowcolor[HTML]{F3F3F3} 
        Conditional Breakpoint & 9.0 \\
        \rowcolor[HTML]{FFFFFF} 
        Tracepoint & 8.6 \\
        \rowcolor[HTML]{F3F3F3} 
        Thread Breakpoint & 8.2 \\
        \rowcolor[HTML]{FFFFFF} 
        Event Breakpoint & 8.2 \\
        \rowcolor[HTML]{F3F3F3} 
        Watchpoint & 8.0 
        \\\hline
        \multicolumn{2}{l}{$^{\mathrm{a}}$Score represents the average degree of developers agreement}
        \\
        \multicolumn{2}{l}{$^{}$for each of the definitions. From 1 (disagree) to 10 (agree).}
        \end{tabular}
    }
    \label{tab:tabDescriptionScore}
\end{table}

\section{Breakpoint Taxonomy}
\label{sec:breakpointTaxonomy}

\begin{figure}[ht]
\centering
\includegraphics[width=\linewidth]{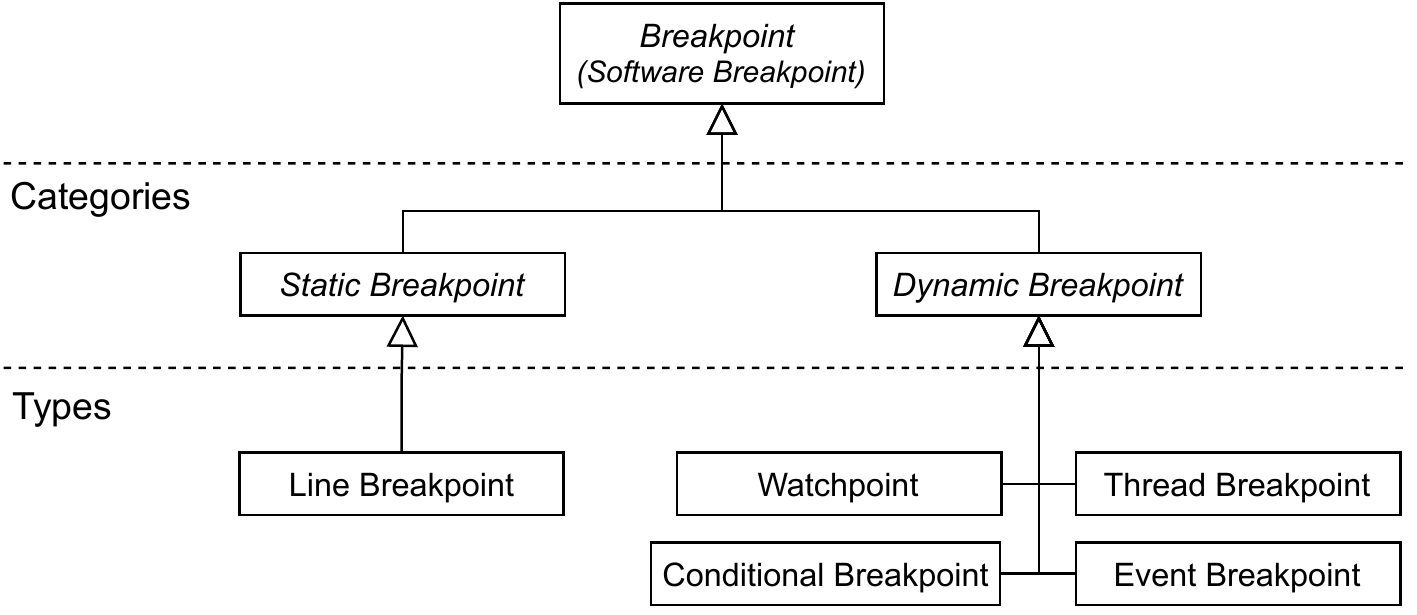}
\caption{\textit{Taxonomy} -- The main breakpoint types organized in two categories.}
\label{fig:figTaxonomy}
\end{figure}

In this section, we present the taxonomy obtained from the mapping. This taxonomy is formed by two major breakpoint categories: Static Breakpoint and Dynamic Breakpoint. In the static breakpoint category, we have the type Line Breakpoint. In the dynamic breakpoit, we have four types: Watchpoint, Condional Breakpoint, Thread Breakpoint, and Event Breakpoint. The Figure \ref{fig:figTaxonomy} show the taxonomy diagram.

The first level of the Taxonomy is the Breakpoint. It is a generalization that includes software-related breakpoints, that is, those that are used in debuggers through IDEs for high-level software and that are not used in debugging directly on hardware, \ie low-level software. We also included the term Software Breakpoint in the diagram as a synonym for Breakpoint, because it is often referred to in academic literature as Software Breakpoint. In addition, the taxonomy does not include Tracepoint type, because this type does not cause the system execution to stop.

The second level of the Taxonomy are the categories Static Breakpoint and Dynamic Breakpoint. We extended these categories from academic literature. Static Breakpoint encompasses all breakpoint types that trigger when the line of code at which the breakpoint is located is the next line to be executed and do not depend on the evaluation of logical expressions or extra events.
Dynamic Breakpoint includes all breakpoint types whose triggering does not depend only or necessarily on the execution of the line of code in which the breakpoint is located, but on some extra event or evaluation of some logical expression. For example, to trigger a Dynamic Breakpoint, it is necessary the condition assigned to the breakpoint is true, or the number of line executions reach a certain number, or call of a certain method, or an object of a certain type is instantiated, and others.

The third level of the taxonomy is composed by breakpoint types. The most often breakpoint type according to the grouping shown in Figure \ref{fig:figTermOccurrences} and Table \ref{tab:tabTermOccurrences} is Line Breakpoint. Furthermore, Line Breakpoint was cited in 23 of the 29 responses, according to the questionnaire. The second term with similar characteristics is Code Breakpoint, which can be attributed as a synonym for Line Breakpoint.

Conditional Breakpoint and Watchpoint are also find in the third level of Taxonomy, in the Dynamic Breakpoint category. Conditional Breakpoint is similar to the Line Breakpoint, but with associated triggering conditions. These conditions are logical expressions that trigger the breakpoint when the line of code is the next line to be executed and the evaluated expression returns true value. Watchpoint is a breakpoint type related to data change, either directly in memory or object instance. For the Watchpoint to be triggered, the data associated with a variable, a field or a memory address must be read, written or changed.

Finally, we describe the Thread Breakpoint and Event Breakpoint types also in the third level. Thread Breakpoint has features applied in distributed programming and multithreading while the Event Breakpoint is applied in monothread executions. Both are triggered when an event happens. For example, Thread Breakpoint triggers when the thread with which it is associated starts or stops. On the other hand, Event Breakpoint is triggered when a specific method or specific function is called, an exception is thrown and others.

\section{Discussion}
\label{sec:discussion}


In this section, we discuss points observed during the analysis by combining the results of RQs 1-3.

From the breakpoint types found in IDEs (RQ1), we recommend that Conditional Breakpoint should not be referred to as a specific breakpoint type as mentioned in one Visual Studio Code IDE documentation. In other IDEs' documentation, there is no term Conditional Breakpoint. Instead, IDEs define conditionals as extra actions that can be applied on which breakpoint type. Through this, any breakpoint type with conditionals is a Dynamic Breakpoint.


Also related to breakpoint type mapping used by gray literature (RQ1), we recommend that Tracepoint should be presented by the IDEs documentation as a feature to support debugging and not as a breakpoint type. The main characteristic of a breakpoint is stop the system running, but Tracepoint and Logpoint do not stop the system execution, just present log messages in the console output.


Analyzing the academic articles (RQ2), we found descriptions of breakpoint types named with specific terms and related to specific technology or specialization, as presented in Table \ref{tab:tabArticlesBreakpoints}. Based on this, we recommend that the more specific is the breakpoint type, the more specific the name given in the literature, thus distancing it from the terminology used by the IDEs documentation and from the terminology known by software developers.


Considering the results of RQs 1-3, we recommend that the suitable term to be used for represent data-related breakpoint type is Data Breakpoint. On academic articles, some authors use the terms Watchpoint and Data Breakpoint as synonyms. Furthermore, the questionnaire revealed that developers confuse the term Watchpoint with an in-memory value monitoring functionality, also known as \textit{Watch windows} or \textit{Watches}, or confuse it with a message logging functionality, such as the Tracepoint breakpoint type.


Observing the terms in the Table \ref{tab:tabBreakpointTermsInIDEs} - RQ1, we identified the same set of characteristics for different terms from different IDEs. For example Inline Breakpoint and Variable Breakpoint are equivalent to Line Breakpoint and Field Watchpoint. We suggest that companies should not create new terms to define their own breakpoint types.


Analysing the questionnaire (RQ3), we found some suggestions of usage of source code instructions as breakpoint type, as \textit{debugger} and \textit{System.Diagnostic.Debugger.Break()}. In addition, the questionnaire revealed that practitioners understand Code Breakpoint as a instruction breakpoint type. Nonetheless, in academic literature we found Code Breakpoint as a generic term to describe Line Breakpoint. We suggest academic literature to adopt the usage of Code Breakpoint as source code instruction.



\section{Threats to Validity}
\label{sec:threatsToValidity}

In this study, we considered the main IDEs used in the software industry and the articles from the academic literature. We have not reviewed all possible IDEs. In this way, this study is not an exhaustive research, and we may have not included some breakpoint types, mainly those of particular technologies or proprietary technologies. Besides that, we searched only documentation related to IDE debuggers, we did not include descriptions of breakpoint types from documentation related to standalone debuggers. 

In our mapping, we only considered breakpoints related to high-level software development oriented IDEs. However, we found in academic literature descriptions of breakpoint types aimed at debugging directly on hardware, \ie low-level software.


\section{Conclusion}
\label{sec:conclusionsFurueWork}

In this article, we present a mapping of the breakpoint types at the software level among different IDEs. We mapped the gray literature on the documentation of the nine main IDEs used by developers according to the three public rankings. In addition, we performed a systematic mapping of academic literature over 68 articles describing breakpoint types. Finally, we analyzed the developers understanding of the main breakpoint types through a questionnaire.

We present three main contributions: 
(1) the mapping of the breakpoint types in the main IDEs and academic literature,
(2) compiled definitions of breakpoint types, and
(3) a breakpoint taxonomy. 
As results, we built a breakpoint type mapping table of the main IDEs, as well as a breakpoint type mapping table for academic papers. Furthermore, we present breakpoint type definitions structured according to attributes mapped in The Grounded Theory method. Finally, we present a taxonomy diagram with two breakpoint type categories. We also bring the discussion about terms used, such as Conditional Breakpoint and Tracepoint, should not necessarily be presented as breakpoint types on IDEs documentation. With the results, we were able to describe the breakpoint types found in the IDEs (RQ1), in the academic literature (RQ2), and by the practitioners (RQ3).

Our breakpoint definitions provide to developers a common lexicon to IDEs. In addition, 
our study contributes to avoid misleading breakpoint names among academic literature and IDE documentation. Furthermore, as suggested by Aniche \al \cite{aniche2021developers}, our taxonomy and definitions can support future debugging research.

As future work, we plan to perform an controlled experiment to evaluate our definitions and study breakpoint types on hardware level. Furthermore, we plan a study on breakpoint types from documentation related to standalone debuggers.

\bibliographystyle{IEEEtran}
\bibliography{references}

\begin{thebibliography}{10}
\providecommand{\url}[1]{#1}
\csname url@samestyle\endcsname
\providecommand{\newblock}{\relax}
\providecommand{\bibinfo}[2]{#2}
\providecommand{\BIBentrySTDinterwordspacing}{\spaceskip=0pt\relax}
\providecommand{\BIBentryALTinterwordstretchfactor}{4}
\providecommand{\BIBentryALTinterwordspacing}{\spaceskip=\fontdimen2\font plus
\BIBentryALTinterwordstretchfactor\fontdimen3\font minus
  \fontdimen4\font\relax}
\providecommand{\BIBforeignlanguage}[2]{{%
\expandafter\ifx\csname l@#1\endcsname\relax
\typeout{** WARNING: IEEEtran.bst: No hyphenation pattern has been}%
\typeout{** loaded for the language `#1'. Using the pattern for}%
\typeout{** the default language instead.}%
\else
\language=\csname l@#1\endcsname
\fi
#2}}
\providecommand{\BIBdecl}{\relax}
\BIBdecl

\bibitem{tanenbaum1973people}
A.~S. Tanenbaum and W.~H. Benson, ``The people's time sharing system,''
  \emph{Software: Practice and Experience}, vol.~3, no.~2, pp. 109--119, 1973.

\bibitem{petrillodevelopers2017}
F.~Petrillo, H.~Mandian, A.~Yamashita, F.~Khomh, and Y.-G. Guéhéneuc, ``How
  do developers toggle breakpoints? observational studies,'' in \emph{2017 IEEE
  International Conference on Software Quality, Reliability and Security
  (QRS)}, 2017, pp. 285--295.

\bibitem{beller2017developers}
M.~Beller, N.~Spruit, and A.~Zaidman, ``How developers debug,'' \emph{PeerJ
  Preprints}, vol.~5, p. e2743v1, 2017.

\bibitem{EclipseDoucmentation}
\BIBentryALTinterwordspacing
``Eclipse documentation.'' [Online]. Available:
  \url{https://help.eclipse.org/\\2021-03/index.jsp}
\BIBentrySTDinterwordspacing

\bibitem{EclipseDoucmentationNews}
\BIBentryALTinterwordspacing
``Debugging the eclipse ide for java developers.'' [Online]. Available:
  \url{https://www.eclipse.org/community/eclipse\_newsletter/2017/june/articl\\e1.php}
\BIBentrySTDinterwordspacing

\bibitem{IntelliJDocumentation}
\BIBentryALTinterwordspacing
``Intellij idea documentation.'' [Online]. Available:
  \url{https://www.jetbrains\\.com/help/idea/using-breakpoints.html\#set-breakpoints}
\BIBentrySTDinterwordspacing

\bibitem{VSDocumentation}
\BIBentryALTinterwordspacing
``Use breakpoints in the visual studio debugger.'' [Online]. Available:
  \url{https://docs.micros\\oft.com/en-us/visualstudio/debugger/using-breakpoints?view=vs-2019}
\BIBentrySTDinterwordspacing

\bibitem{VSCodeDocumentation}
\BIBentryALTinterwordspacing
``User guide - debugging.'' [Online]. Available:
  \url{https://code.v\\isualstudio.com/docs/editor/debugging}
\BIBentrySTDinterwordspacing

\bibitem{DiffWatchBreakQuora}
\BIBentryALTinterwordspacing
``What is the difference between watchpoints and breakpoints in eclipse
  debugging?'' [Online]. Available:
  \url{https://www.quora.com/What-is-the-difference-between-watchpoints-and-breakpoints-in-Eclipse-debugging}
\BIBentrySTDinterwordspacing

\bibitem{DiffWatchMethodStack}
\BIBentryALTinterwordspacing
``Difference between breakpoint on method signature vs breakpoint on first line
  in method.'' [Online]. Available:
  \url{https://stackoverflow.com/que\\stions/26364634/difference-between-breakpoint-on-method-signature-vs-breakpoint-on-first-line-in}
\BIBentrySTDinterwordspacing

\bibitem{DiffBreakWatchSAP}
\BIBentryALTinterwordspacing
``Difference between watch point and break point.'' [Online]. Available:
  \url{https://answers.sap.com/questions/4181705/difference-between-watch-\\point-and-break-point.html}
\BIBentrySTDinterwordspacing

\bibitem{DiffBreakWatchLessBro}
\BIBentryALTinterwordspacing
``Difference between watchpoint and breakpoint?'' [Online]. Available:
  \url{https://www.lessbro.com/Difference-between-watchpoint-and-breakpoi\\nt-/49}
\BIBentrySTDinterwordspacing

\bibitem{DiffStaticDynamicGoCoding}
\BIBentryALTinterwordspacing
``Difference between a static breakpoint and dynamic breakpoint.'' [Online].
  Available:
  \url{https://gocoding.org/difference-between-a-static-br\\eakpoint-and-dynamic-breakpoint/}
\BIBentrySTDinterwordspacing

\bibitem{ConditionalBreakVSMagazine}
\BIBentryALTinterwordspacing
``Understand conditional breakpoints in c++.'' [Online]. Available:
  \url{https://visualstudiomagazine.com/articles/2016/09/01/understand-condit\\ional-breakpoints.aspx}
\BIBentrySTDinterwordspacing

\bibitem{DataBreakVS2019MicrosoftDevBlog}
\BIBentryALTinterwordspacing
``Break when value changes: Data breakpoints for .net core in visual studio
  2019.'' [Online]. Available:
  \url{https://devblogs.microsoft.com/visual\\studio/break-when-value-changes-data-breakpoints-for-net-core-in-visua\\l-studio-2019/}
\BIBentrySTDinterwordspacing

\bibitem{DataBreakVS2017MicrosoftDevBlog}
\BIBentryALTinterwordspacing
``Break when value changes: Data breakpoints for .net core in visual studio
  2019.'' [Online]. Available: \url{Data Breakpoints – Visual Studio 2017
  15.8 Update}
\BIBentrySTDinterwordspacing

\bibitem{petrillo2019swarm}
F.~Petrillo, Y.-G. Gu{\'e}h{\'e}neuc, M.~Pimenta, C.~D.~S. Freitas, and
  F.~Khomh, ``Swarm debugging: The collective intelligence on interactive
  debugging,'' \emph{Journal of Systems and Software}, vol. 153, pp. 152--174,
  2019.

\bibitem{johnson1982software}
M.~S. Johnson, ``A software debugging glossary,'' \emph{ACM Sigplan Notices},
  vol.~17, no.~2, pp. 53--70, 1982.

\bibitem{keppel1993fast}
D.~Keppel, ``Fast data breakpoints,'' 1993.

\bibitem{vasudevan2009re}
A.~Vasudevan, ``Re-inforced stealth breakpoints,'' in \emph{2009 Fourth
  International Conference on Risks and Security of Internet and Systems
  (CRiSIS 2009)}.\hskip 1em plus 0.5em minus 0.4em\relax IEEE, 2009, pp.
  59--66.

\bibitem{spinellis2006debuggers}
D.~Spinellis, ``Debuggers and logging frameworks,'' \emph{IEEE software},
  vol.~23, no.~3, pp. 98--99, 2006.

\bibitem{arya2017transition}
K.~Arya, T.~Denniston, A.~Rabkin, and G.~Cooperman, ``Transition watchpoints:
  Teaching old debuggers new tricks,'' \emph{arXiv preprint arXiv:1703.10864},
  2017.

\bibitem{kumar2013behave}
A.~Kumar, P.~Goodman, A.~Goel, and A.~D. Brown, ``Behave or be watched:
  Debugging with behavioral watchpoints,'' in \emph{Proceedings of the 9th
  Workshop on Hot Topics in Dependable Systems}, 2013, pp. 1--6.

\bibitem{copperman1995poor}
M.~Copperman and J.~Thomas, ``Poor man's watchpoints,'' \emph{ACM SIGPLAN
  Notices}, vol.~30, no.~1, pp. 37--44, 1995.

\bibitem{zhao2008million}
Q.~Zhao, R.~Rabbah, S.~Amarasinghe, L.~Rudolph, and W.-F. Wong, ``How to do a
  million watchpoints: Efficient debugging using dynamic instrumentation,'' in
  \emph{International Conference on Compiler Construction}.\hskip 1em plus
  0.5em minus 0.4em\relax Springer, 2008, pp. 147--162.

\bibitem{chew2010kivati}
L.~Chew and D.~Lie, ``Kivati: fast detection and prevention of atomicity
  violations,'' in \emph{Proceedings of the 5th European conference on Computer
  systems}, 2010, pp. 307--320.

\bibitem{sommer2013minerva}
P.~Sommer and B.~Kusy, ``Minerva: Distributed tracing and debugging in wireless
  sensor networks,'' in \emph{Proceedings of the 11th ACM Conference on
  Embedded Networked Sensor Systems}, 2013, pp. 1--14.

\bibitem{jang2019revisiting}
J.~Jang and B.~B. Kang, ``Revisiting the arm debug facility for os kernel
  security,'' in \emph{2019 56th ACM/IEEE Design Automation Conference
  (DAC)}.\hskip 1em plus 0.5em minus 0.4em\relax IEEE, 2019, pp. 1--6.

\bibitem{aniche2021developers}
M.~Aniche, C.~Treude, and A.~Zaidman, ``How developers engineer test cases: An
  observational study,'' \emph{arXiv preprint arXiv:2103.01783}, 2021.

\bibitem{stackOverflowSurvey2019}
\BIBentryALTinterwordspacing
``Stack overflow developer survey 2019.'' [Online]. Available:
  \url{https://in\\sights.stackoverflow.com/survey/2019}
\BIBentrySTDinterwordspacing

\bibitem{googleTrends2020}
\BIBentryALTinterwordspacing
``Google trends 2020.'' [Online]. Available:
  \url{https://pypl.github.io/IDE.h\\tml}
\BIBentrySTDinterwordspacing

\bibitem{g22020}
\BIBentryALTinterwordspacing
``Best integrated development environments (ide).'' [Online]. Available:
  \url{https://www.g2.com/categories/integra\\ted-development-environments-ide}
\BIBentrySTDinterwordspacing

\bibitem{glaser2017discovery}
B.~G. Glaser and A.~L. Strauss, \emph{Discovery of grounded theory: Strategies
  for qualitative research}.\hskip 1em plus 0.5em minus 0.4em\relax Routledge,
  2017.

\bibitem{haigh2016eniac}
T.~Haigh, P.~M. Priestley, M.~Priestley, and C.~Rope, \emph{ENIAC in action:
  Making and remaking the modern computer}.\hskip 1em plus 0.5em minus
  0.4em\relax MIT press, 2016.

\bibitem{haigh2014engineering}
T.~Haigh, M.~Priestley, and C.~Rope, ``Engineering" the miracle of the eniac":
  Implementing the modern code paradigm,'' \emph{IEEE Annals of the History of
  Computing}, vol.~36, no.~2, pp. 41--59, 2014.

\bibitem{tropp_holberton_1973}
\BIBentryALTinterwordspacing
H.~S. Tropp and F.~E. Holberton, ``Computer oral history collection, 1969-1973,
  1977,'' p. 56–57, Apr 1973. [Online]. Available:
  \url{https://amhi\\story.si.edu/archives/AC0196\_bart730427.pdf}
\BIBentrySTDinterwordspacing

\bibitem{cristobal2011ibm}
\BIBentryALTinterwordspacing
B.~Cristobal, \emph{IBM Oliver (CICS Interactive Test/Debug)}.\hskip 1em plus
  0.5em minus 0.4em\relax Cede Publishing, 2011. [Online]. Available:
  \url{https://books.google.ca/books?\\id=JbkeygAACAAJ}
\BIBentrySTDinterwordspacing

\bibitem{carroll1985programming}
\BIBentryALTinterwordspacing
D.~Carroll, \emph{Programming with Turbo Pascal}, ser. Byte books.\hskip 1em
  plus 0.5em minus 0.4em\relax Micro Text Productions, 1985, no. v. 1.
  [Online]. Available: \url{https://books.google.ca/books?id=jxESAQAAMAAJ}
\BIBentrySTDinterwordspacing

\bibitem{StallmanR.Pesch2002}
R.~Stallman and S.~Shebs, \emph{{Debugging with GDB - The GNU Source-Level
  Debugger}}.\hskip 1em plus 0.5em minus 0.4em\relax GNU Press, 2002.

\bibitem{vostokov2008windbg}
D.~Vostokov, \emph{WinDbg: A Reference Poster and Learning Cards}.\hskip 1em
  plus 0.5em minus 0.4em\relax OpenTask, 2008.

\bibitem{Chern:2007}
\BIBentryALTinterwordspacing
R.~Chern and K.~De~Volder, ``Debugging with control-flow breakpoints,'' in
  \emph{Proceedings of the 6th International Conference on Aspect-oriented
  Software Development}, ser. AOSD '07.\hskip 1em plus 0.5em minus 0.4em\relax
  New York, NY, USA: ACM, 2007, pp. 96--106. [Online]. Available:
  \url{http://doi.acm.org/10.1145/1218563.1218575}
\BIBentrySTDinterwordspacing

\bibitem{beller2018dichotomy}
M.~Beller, N.~Spruit, D.~Spinellis, and A.~Zaidman, ``On the dichotomy of
  debugging behavior among programmers,'' in \emph{40th International
  Conference on So ware Engineering, ICSE}, 2018, pp. 572--583.

\bibitem{wahbe1992efficient}
R.~Wahbe, ``Efficient data breakpoints,'' \emph{ACM SIGPLAN Notices}, vol.~27,
  no.~9, pp. 200--212, 1992.

\bibitem{dupriez2017analysis}
T.~Dupriez, G.~Polito, and S.~Ducasse, ``Analysis and exploration for new
  generation debuggers,'' in \emph{Proceedings of the 12th edition of the
  International Workshop on Smalltalk Technologies}, 2017, pp. 1--6.

\bibitem{corrodi2016towards}
C.~Corrodi, ``Towards efficient object-centric debugging with declarative
  breakpoints.'' in \emph{SATToSE}, 2016, pp. 32--39.

\bibitem{charmaz2006constructing}
K.~Charmaz, \emph{Constructing grounded theory: A practical guide through
  qualitative analysis}.\hskip 1em plus 0.5em minus 0.4em\relax sage, 2006.

\bibitem{boix2012handling}
E.~G. Boix, ``Handling partial failures in mobile ad hoc network applications:
  From programming language design to tool support,'' \emph{Vrije Universiteit
  Brussel, Faculty of Sciences, Software Languages Lab (Ph. D. Thesis)}, 2012.

\bibitem{lencevicius1999query}
R.~Lencevicius, \emph{Query-based debugging}.\hskip 1em plus 0.5em minus
  0.4em\relax University of California, Santa Barbara, 1999.

\bibitem{maruyama2003debugging}
K.~Maruyama and M.~Terada, ``Debugging with reverse watchpoint,'' in
  \emph{Third International Conference on Quality Software, 2003.
  Proceedings.}\hskip 1em plus 0.5em minus 0.4em\relax IEEE, 2003, pp.
  116--123.

\bibitem{visan2012temporal}
A.-M. Visan, ``Temporal meta-programming: treating time as a spatial
  dimension,'' Ph.D. dissertation, Northeastern University, 2012.

\bibitem{mcdowell1989debugging}
C.~E. McDowell and D.~P. Helmbold, ``Debugging concurrent programs,'' \emph{ACM
  Computing Surveys (CSUR)}, vol.~21, no.~4, pp. 593--622, 1989.

\bibitem{jahne1996peard}
A.~J{\"a}hne, S.~D. Urban, and S.~W. Dietrich, ``Peard: A prototype environment
  for active rule debugging,'' \emph{Journal of Intelligent Information
  Systems}, vol.~7, no.~2, pp. 111--128, 1996.

\bibitem{kosar2014debugging}
T.~Kosar, M.~Mernik, J.~Gray, and T.~Kos, ``Debugging measurement systems using
  a domain-specific modeling language,'' \emph{Computers in industry}, vol.~65,
  no.~4, pp. 622--635, 2014.

\bibitem{seaton2014debugging}
C.~Seaton, M.~L. Van De~Vanter, and M.~Haupt, ``Debugging at full speed,'' in
  \emph{Proceedings of the Workshop on Dynamic Languages and Applications},
  2014, pp. 1--13.

\bibitem{alsallakh2012visual}
B.~Alsallakh, P.~Bodesinsky, A.~Gruber, and S.~Miksch, ``Visual tracing for the
  eclipse java debugger,'' in \emph{2012 16th European Conference on Software
  Maintenance and Reengineering}.\hskip 1em plus 0.5em minus 0.4em\relax IEEE,
  2012, pp. 545--548.

\bibitem{kumar2020amber}
A.~Kumar, Z.~Wang, S.~Ni, and C.~Li, ``Amber: a debuggable dataflow system
  based on the actor model,'' \emph{Proceedings of the VLDB Endowment},
  vol.~13, no.~5, pp. 740--753, 2020.

\bibitem{andreetta_fontana_eduardo_2021_5213971}
\BIBentryALTinterwordspacing
E.~Andreetta~Fontana and F.~Petrillo, ``Breakpoints into the wild: an
  exploratory study,'' Aug. 2021. [Online]. Available:
  \url{https://doi.org/10.5281/zenodo.5663903}
\BIBentrySTDinterwordspacing

\end{thebibliography}

\end{document}